\documentclass[twocolumn,amsmath,amssymb]{revtex4-1}
\usepackage{graphicx}
\usepackage{amssymb}
\usepackage{amsmath}
\usepackage{subfigure}
\usepackage{graphpap}
\usepackage{dcolumn} 
\usepackage{bm}
\usepackage{color,calc}
\usepackage{longtable}
\usepackage{color}
\usepackage{tikz}

\usepackage{multirow}
\usepackage{hyperref}

\usepackage{txfonts}

\newcommand{\Eq}[1]{Eq.~(\ref{#1})}

\newcommand{\Refs}[1]{Refs. \onlinecite{#1}}
\newcommand{\Ref}[1]{Ref. \onlinecite{#1}}
\newcommand{\Fig}[1]{Fig. \ref{#1}}
\newcommand{\Tab}[1]{Table \ref{#1}}
\newcommand{\Sec}[1]{Sec. \ref{#1}}

\newcommand{\R}{\mathbf{r}}

\newcommand{\funcder}[2]{\frac{\delta #1}{\delta #2}}

\newcommand{\lr}{\ensuremath{\text{lr}}}

\newcommand{\bra}[1]{\ensuremath{\langle #1 \vert}}
\newcommand{\ket}[1]{\ensuremath{\vert #1  \rangle}}
\newcommand{\braket}[2]{\ensuremath{\langle  #1 \vert #2  \rangle}}

\renewcommand{\b}[1]{\ensuremath{\mathbf{#1}}}

\renewcommand{\b}[1]{\ensuremath{\mathbf{#1}}}

\renewcommand{\d}{\text{d}}
\begin{document}

\title{Self-consistent range-separated density-functional theory with second-order perturbative correction via the optimized-effective-potential method}
\author{Szymon \'Smiga$^{1}$}\email{szsmiga@fizyka.umk.pl}
\author{Ireneusz Grabowski$^{1}$}\email{ig@fizyka.umk.pl}
\author{Mateusz Witkowski$^{1}$}
\author{Bastien Mussard$^{2}$}
\author{Julien Toulouse$^{3}$}\email{toulouse@lct.jussieu.fr}
\affiliation{
$^1$Institute of Physics, Faculty of Physics, Astronomy and Informatics, Nicolaus Copernicus University, 87-100 Toru\'n, Poland\\
$^2$Department of Chemistry and Biochemistry, University of Colorado Boulder, CO 80302 Boulder, USA\\
$^3$Laboratoire de Chimie Th\'eorique (LCT), Sorbonne Universit\'e and CNRS, F-75005 Paris, France\\
}

\begin{abstract}
We extend the range-separated double-hybrid RSH+MP2 method [J. G. \'Angy\'an \textit{et al.}, Phys. Rev. A {\bf 72}, 012510 (2005)], combining long-range HF exchange and MP2 correlation with a short-range density functional, to a fully self-consistent version using the optimized-effective-potential technique in which the orbitals are obtained from a local potential including the long-range HF and MP2 contributions. We test this approach, that we name RS-OEP2, on a set of small closed-shell atoms and molecules. For the commonly used value of the range-separation parameter $\mu=0.5$ bohr$^{-1}$, we find that self-consistency does not seem to bring any improvement for total energies, ionization potentials, and electronic affinities. However, contrary to the non-self-consistent RSH+MP2 method, the present RS-OEP2 method gives a LUMO energy which physically corresponds to a neutral excitation energy and gives local exchange-correlation potentials which are reasonably good approximations to the corresponding Kohn-Sham quantities. At a finer scale, we find that RS-OEP2 gives largely inaccurate correlation potentials and correlated densities, which points to the need of further improvement of this type of range-separated double hybrids.
\end{abstract}

\date{December 3, 2019}

\maketitle

\section{Introduction}

Kohn-Sham (KS) density-functional theory (DFT)~\cite{HohKoh-PR-64,KohSha-PR-65} is among the most popular approaches to describe properties of atoms, molecules, and solids. The key quantity in KS DFT is the exchange-correlation density functional which describes the non-classical part of the electron-electron interaction. Despite the enormous effort put in studying the exchange-correlation density functional, its exact explicit form still remains unknown. To date, however, many useful density-functional approximations (DFAs) have been proposed. They are commonly classified by increasing complexity according to Perdew's ladder of DFAs~\cite{pardew:2001:ladder}: the local-density approximation (LDA), semilocal generalized-gradient approximations (GGAs) and meta-GGA approximations, hybrid approximations including a fraction of Hartree-Fock (HF) exchange~\cite{Bec-JCP-93}, and more sophisticated approximations including a dependence on orbital energies and virtual orbitals such as double-hybrid approximations~\cite{Gri-JCP-06,ShaTouSav-JCP-11,SmiFraMusBukGraLupTou-JCP-16}, random-phase approximations (RPAs)~\cite{Fur-PRB-01,AngLiuTouJan-JCTC-11,rpa}, the so-called \textit{ab initio} DFT~\cite{grabowski:2007:ccpt2,verma:044105,grabowski13,grabowski:2014:jcp,SCSIP}, and interaction-strength interpolation (ISI) approximations~\cite{SeiPerKur-PRL-00,Edo-ISI-2019}. 

The last family of approximations also includes range-separated hybrid (RSH) wave-function/density-functional approaches~\cite{Sav-INC-96,TouColSav-PRA-04}, in which the Coulomb electron-electron interaction $w_\text{ee}(r_{12}) =1/r_{12}$ is decomposed as 
\begin{eqnarray}\label{intmu}
w_\text{ee}(r_{12}) = w_\text{ee}^{\text{lr}}(r_{12}) + w_\text{ee}^{\text{sr}}(r_{12}),
\end{eqnarray}
where $w_\text{ee}^{\text{lr}}(r_{12})=\text{erf}(\mu r_{12})/r_{12}$ is a long-range interaction (written with the error function erf), $w_\text{ee}^{\text{sr}}(r_{12})=\text{erfc}(\mu r_{12})/r_{12}$ is the complementary short-range interaction (written with the complementary error function erfc), and $\mu$ is the range-separation parameter which roughly represents the inverse of the distance where the short-range interaction transits to the long-range one. In these approaches, the long-range contribution is described by a wave-function method, e.g. second-order M{\o}ller-Plesset (MP2) perturbation theory (resulting in a method called RSH+MP2)~\cite{AngGerSavTou-PRA-05}, RPA (resulting in a method that we will call RSH+RPA)~\cite{TouGerJanSavAng-PRL-09,JanHenScu-JCP-09} , or coupled cluster~\cite{GolWerSto-PCCP-05}, and the complementary short-range contribution is described by specially developed short-range semilocal DFAs~\cite{TouSavFla-IJQC-04,TouColSav-PRA-04,TouColSav-JCP-05,PazMorGorBac-PRB-06,GolWerStoLeiGorSav-CP-06}.

In the previously listed range-separated approaches, the orbitals and orbital energies are calculated within the generalized Kohn-Sham (GKS) scheme~\cite{SeiGorVogMajLev-PRB-96} where KS-like equations are solved with a nonlocal long-range HF exchange potential, but without adding to the potential the contribution coming from the long-range wave-function correlation part. The long-range correlation energy is only computed a posteriori using the orbitals and orbital energies from the previously solved GKS equations. The lack of full self-consistency of these approaches may affect their accuracy and complicates the calculation of molecular properties.

Recently, He{\ss}elmann \textit{et al.}~\cite{Hesselmann2018} have reported a fully self-consistent RSH+RPA method based on the optimized-effective-potential (OEP) scheme~\cite{ShaHor-PR-53,TalSha-PRA-76} where the total RSH+RPA energy is computed using the orbitals optimized in the presence of a fully local exchange-correlation potential including the long-range RPA correlation contribution. Tests of this method on molecular reaction energies and properties are promising and encourage us to test a similar fully self-consistent RSH+MP2 method based on the OEP approach. Indeed, the RSH+MP2 method represents the simplest variant of the correlated range-separated approaches and its accuracy has been well investigated~\cite{AngGerSavTou-PRA-05,GerAng-CPL-05b,GolStoThiSch-PRA-07,GolLeiManMitWerSto-PCCP-08,GolWerSto-CP-08,JulNCI2010,JulWI2010,ChaStoWerLei-MP-10,ChaJacAdaStoLei-JCP-10,JulienSolid2015,TayBT2016,Jul2017Frac}, revealing that it is capable of providing a rather significant improvement over standard semilocal DFAs, especially for the description of weak interaction energies in molecules\cite{AngGerSavTou-PRA-05,GerAng-CPL-05b,GolLeiManMitWerSto-PCCP-08,GolWerSto-CP-08,JulNCI2010,JulWI2010,TayBT2016} and solids\cite{JulienSolid2015}, while keeping a reasonably low computational cost for a fifth-rung approximation. Thus, in this work, we have developed the method where the RSH+MP2 energy is optimized with respect to variations of the orbitals and orbital energies via a fully local exchange-correlation potential including the long-range MP2 correlation contribution. We name this method RS-OEP2.
 
There at least two potential advantages for having a fully self-consistent method. First, we expect an improvement in spin-unrestricted calculations for symmetry breaking and open-shell situations, as observed in \Ref{PevHea-JCP-13}. Second, calculations of properties by response theory are facilitated. Moreover, there are at least two motivations for realizing this self-consistency with a fully local potential. First, we stay within the KS scheme which allows us to perform comparisons of local potentials with other DFAs. Second, contrary to the GKS scheme, the virtual orbitals are approximations of neutral excited states and are thus ``ready'' for time-dependent density-functional theory (TDDFT) calculations.

The approach is similar to the self-consistent OEP double-hybrid method based on a linear decomposition of the electron-electron interaction, $w_\text{ee}(r_{12}) = \lambda w_\text{ee}(r_{12}) + (1-\lambda) w_\text{ee}(r_{12})$, named 1DH-OEP, that was developed in \Ref{SmiFraMusBukGraLupTou-JCP-16}.
However, there at least two important advantages of the present RS-OEP2 method over the 1DH-OEP method. First, in the RS-OEP2 method, the whole long-range part of the exchange-correlation potential is treated by OEP, which leads to an exchange-correlation potential correctly decaying as $-1/r$ asymptotically. In comparison, the OEP-1DH method gives an exchange-correlation potential decaying as $-\lambda/r$ and thus underestimates the long-range tail. Second, the RS-OEP2 method will have a fast convergence with the size of the one-electron basis set since the entire short-range correlation part is treated by DFT~\cite{FraMusLupTou-JCP-15}. By contrast, in the OEP-1DH method, a fraction of the short-range correlation is treated in the second-order perturbative term, which slows down the basis convergence.

The papers is organized as follows. In Sec.~\ref{sec:theory}, we briefly review the standard RSH+MP2 method and present its self-consistent extension based on the OEP scheme. After giving computational details in Sec.\ref{sec:comput}, we then discuss in Sec.~\ref{sec:results} the results obtained for a few atoms (He, Be, Ne, Ar) and molecules (CO and H$_2$O) on total energies, ionization potentials (IPs), electronic affinities (EAs), exchange and correlation potentials, and correlated densities. Sec.~\ref{sec:conclusion} contains our conclusions. Throughout the paper, we use the convention that $i$ and $j$ indices label occupied spin orbitals, $a$ and $b$ label virtual ones, and $p$ and $q$ are used for both occupied and virtual spin orbitals. In all equations, Hartree atomic units are assumed.

\section{Theory}
\label{sec:theory}

\subsection{The standard RSH+MP2 method}

In the standard RSH+MP2 method the total energy expression is defined as~\cite{AngGerSavTou-PRA-05}
\begin{eqnarray}
E &=& \sum_i
\int \varphi_i^*({\bf x}) \left( -\frac{1}{2} \bm{\nabla}^2 +v_{\text{ne}}(\b{x}) \right) \varphi_i({\bf x}) \; \d {\bf x}
+ E_\text{H} + E_\text{xc}^\text{RSH+MP2},\;
\nonumber\\
\label{ERSH}
\end{eqnarray}
where $\{\varphi_i({\bf x})\}$ are the occupied spin orbitals written with space-spin coordinates $\b{x}=(\b{r},\sigma)$, $v_{\text{ne}}(\b{r})$ is the nuclei-electron potential, and $E_\text{H} = (1/2) \iint \rho({\bf x}_1) \rho({\bf x}_2) w_\text{ee}(r_{12}) \d {\bf x}_1 \d {\bf x}_2$ is the Hartree energy written with the spin densities $\rho({\bf x}) = \sum_i|\varphi_i({\bf x})|^2$. The exchange-correlation energy $E_\text{xc}^\text{RSH+MP2}$ is defined as
\begin{equation}
E_\text{xc}^\text{RSH+MP2} = E_\text{xc}^\text{sr,DFA} + E_\text{x}^\text{lr,HF} + E_\text{c}^\text{lr,MP2},
\label{ExcRSH}
\end{equation}
where $E_\text{xc}^\text{sr,DFA}$ is a complement short-range exchange-correlation semilocal DFA, and the last two terms correspond to the long-range HF (or exact) exchange energy
\begin{equation}
E_\text{x}^{\lr,\text{HF}} = - \frac{1}{2} \sum_{i,j}
\braket{ij}{ji}^\text{lr},
\label{ExHF}
\end{equation}
and to the long-range MP2 correlation energy 
\begin{equation}
E_\text{c}^{\text{lr,MP2}} = - \frac{1}{4} \sum_{i,j}
\sum_{a,b}
 \frac{|\bra{ij}\ket{ab}^\text{lr}|^2}{\varepsilon_a + \varepsilon_b - \varepsilon_i - \varepsilon_j},
\label{EcMP2}
\end{equation}
where
\begin{equation}
\braket{pq}{rs}^{\lr}=\iint \d {\bf x}_1 \d {\bf x}_2 \varphi_p^*({\bf x}_1) \varphi_q^*({\bf x}_2) w_\text{ee}^\lr(r_{12}) \varphi_r({\bf x}_1) \varphi_s({\bf x}_2) 
\end{equation}
are the long-range two-electron integrals associated with the long-range interaction $w_\text{ee}^\lr(r_{12})$ and  $\varepsilon_p$ are the orbital eigenvalues.

The orbitals are calculated using the GKS scheme~\cite{SeiGorVogMajLev-PRB-96} where we disregard the long-range MP2 correlation contribution, resulting in the so-called RSH equations
\begin{eqnarray}
\left( -\frac{1}{2} \bm{\nabla}^2 +v_{\text{ne}}(\b{r}) + v_\text{H}(\b{r}) +  v_\text{xc}^{\text{sr,DFA}}(\b{x}) \right) \varphi_p({\bf x})  
\nonumber\\
+\int v_{x}^\text{lr,HF}(\b{x},\b{x'}) \varphi_p({\bf x}') \text{d}\b{x}' = \varepsilon_p \varphi_p({\bf x}) \; ,
\label{rhsks}
\end{eqnarray}
where $v_\text{H}(\b{r}) = \int \rho({\bf x}') w_\text{ee}(|\b{r}-\b{r}'|) \d {\bf x}'$ is the Hartree potential, $v_\text{x}^\text{sr,DFA}(\b{x}) = \delta  E_\text{x}^\text{sr,DFA} /\delta \rho({\bf x})$ is the short-range exchange-correlation DFA potential, and $v_{x}^\text{lr,HF}(\b{x},\b{x'}) =- \sum_i \varphi_i^*(\b{x}') \varphi_i(\b{x}) w_\text{ee}^\lr(|\b{r}-\b{r}'|)$ is the long-range nonlocal HF potential. After solving Eq.~(\ref{rhsks}), the calculated orbitals and orbital energies are used in Eq.~(\ref{ERSH}) to obtain the RSH+MP2 energy.

\begin{figure*}
\includegraphics[scale=0.45]{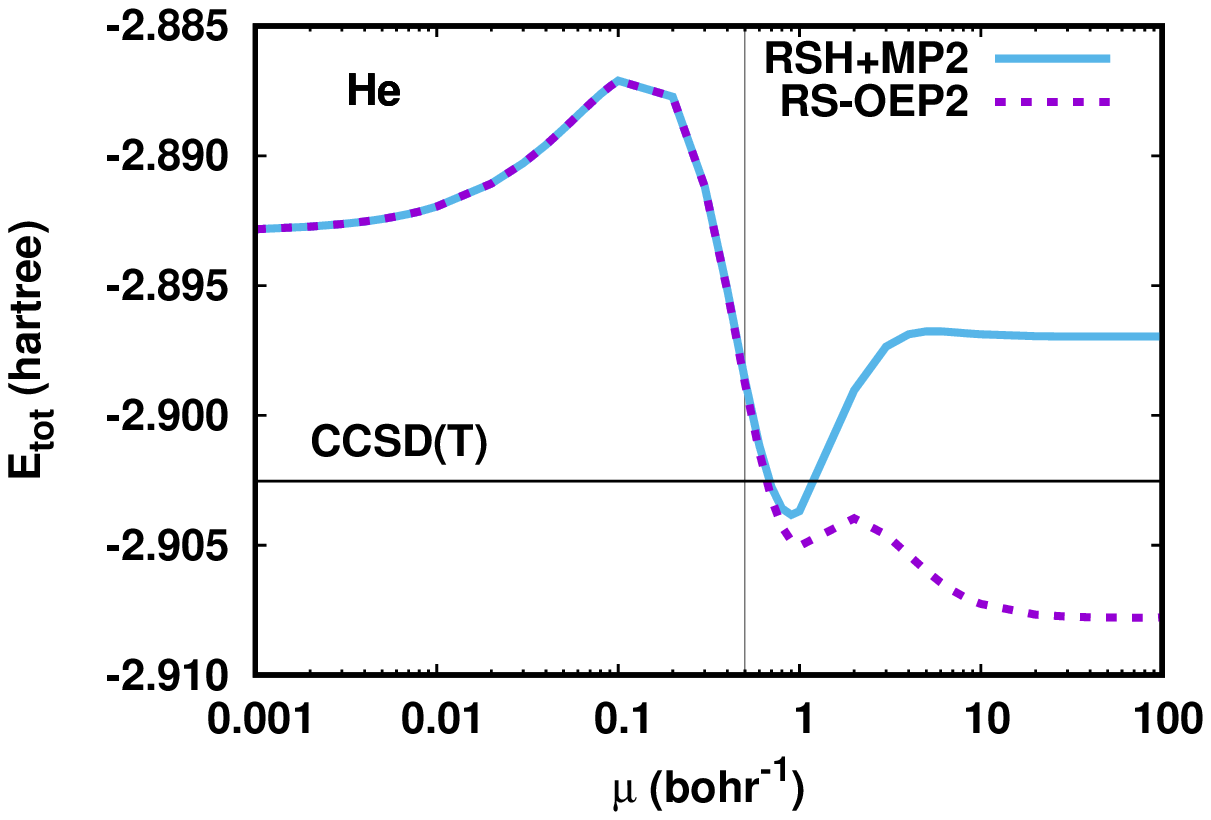}
\includegraphics[scale=0.45]{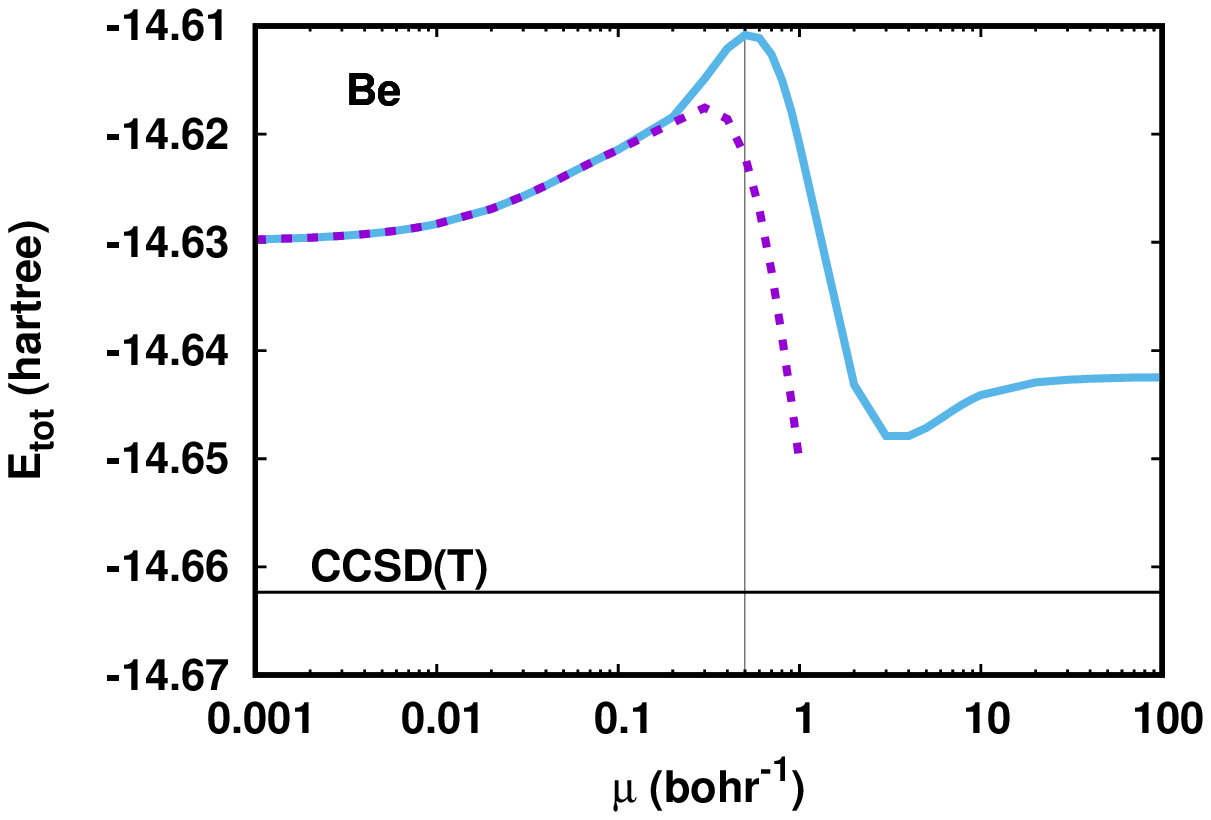}
\includegraphics[scale=0.45]{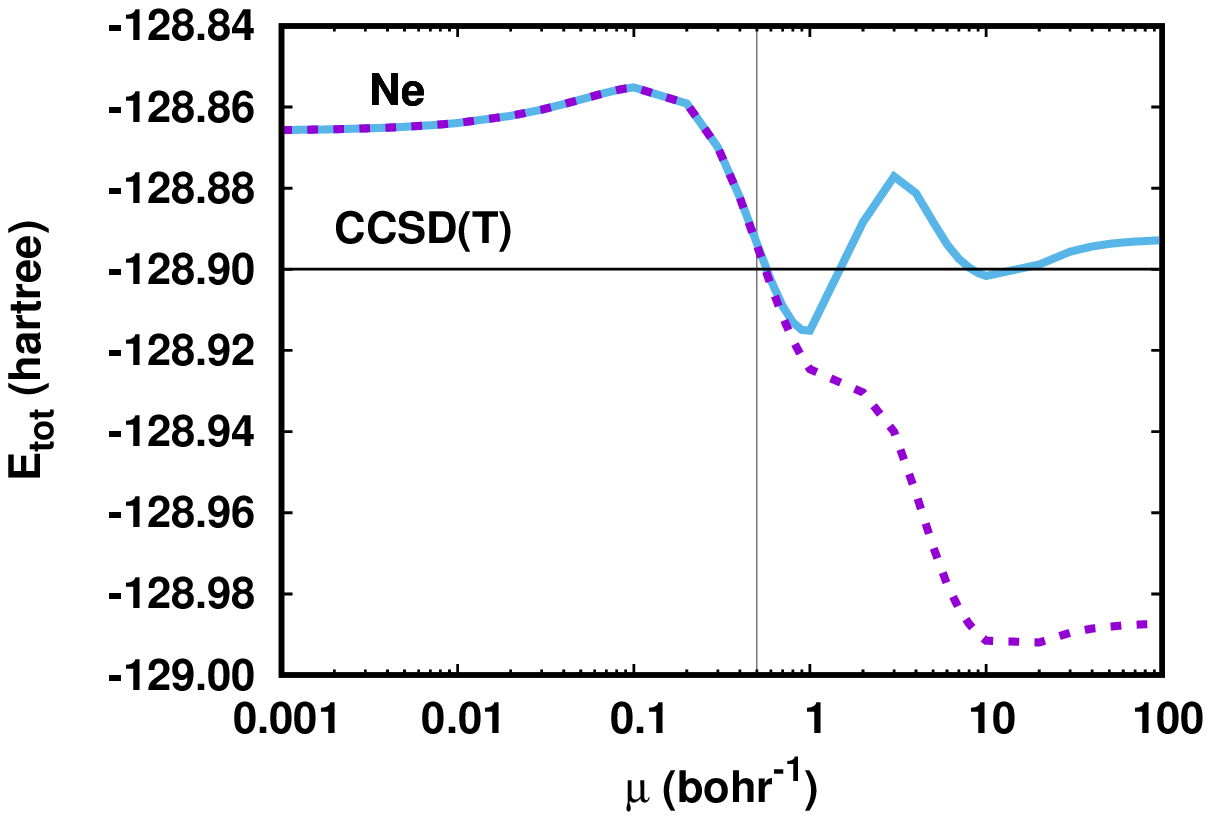}
\includegraphics[scale=0.45]{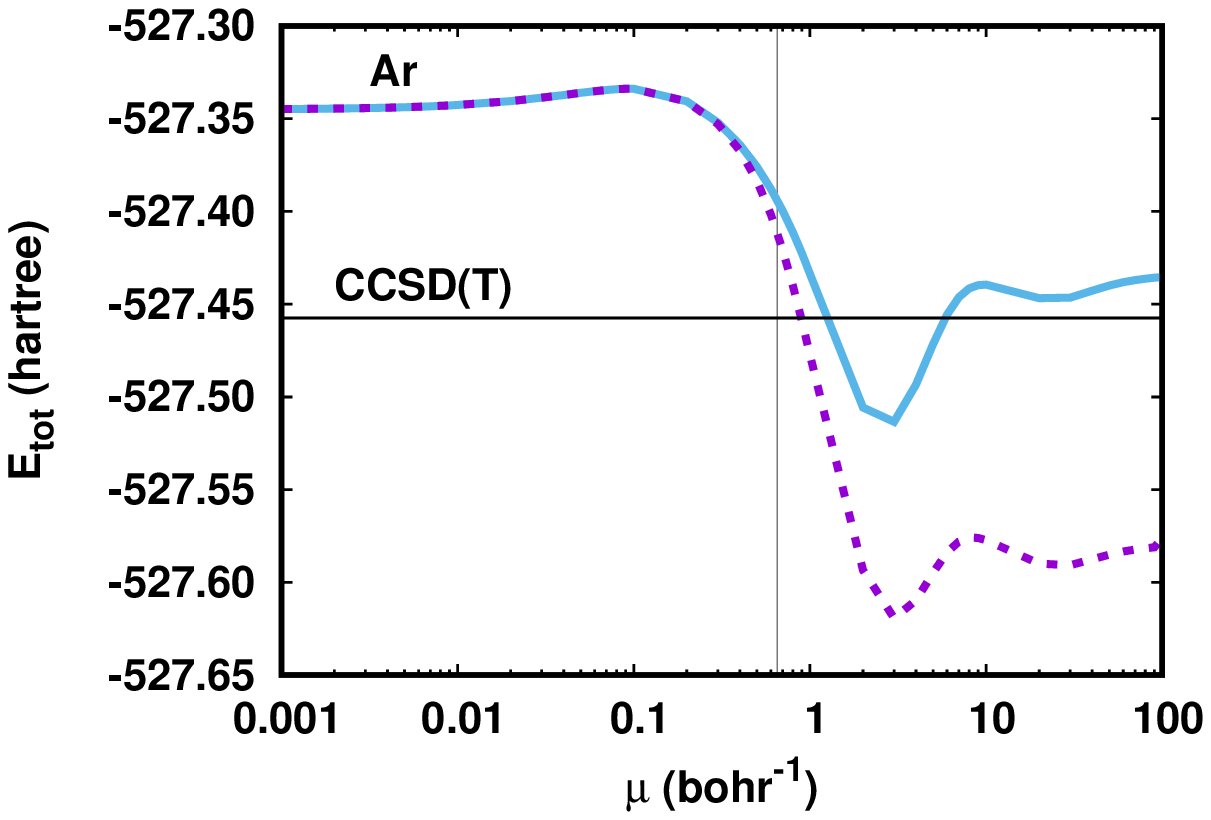}
\includegraphics[scale=0.45]{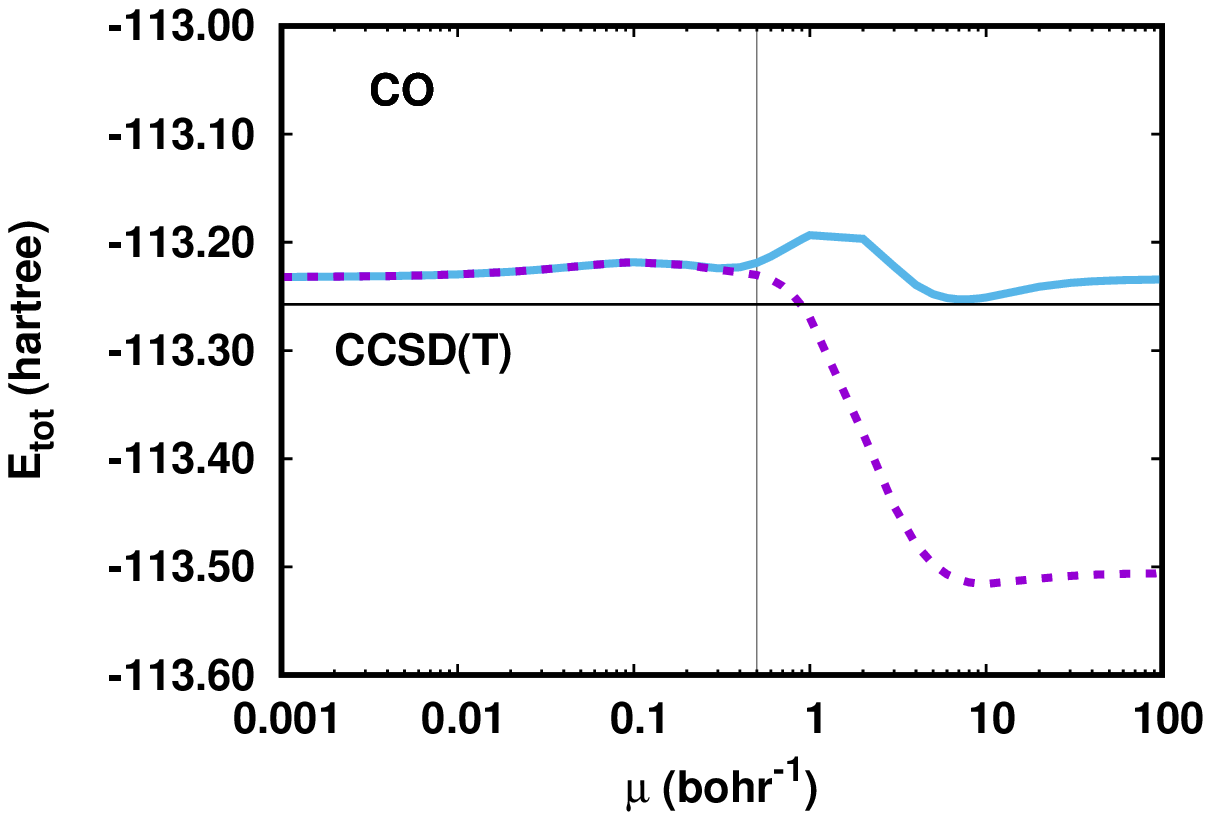}
\includegraphics[scale=0.45]{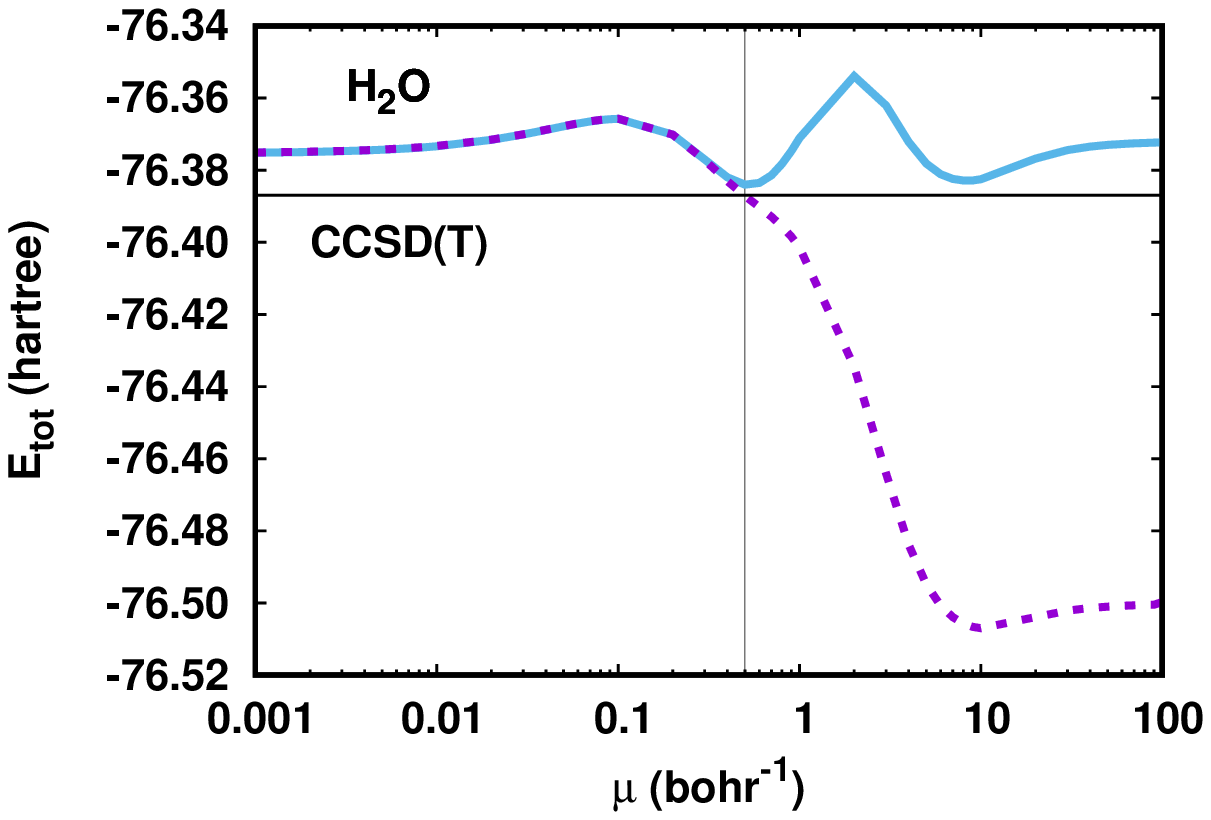}
\caption{Total energies calculated with the RSH+MP2 and RS-OEP2 methods with the srPBE exchange-correlation density functional as a function of the range-separation parameter $\mu$. The reference total energies are calculated with the CCSD(T) method in the same basis set (horizontal black line). The vertical lines correspond to commonly used value $\mu=0.5$ bohr$^{-1}$. For Be, the RS-OEP2 calculations are unstable for $\mu >1.0$ bohr$^{-1}$.
}
\label{fig:etot}
\end{figure*}

\subsection{A self-consistent version of RSH+MP2: The RS-OEP2 method}
Similarly to \Ref{SmiFraMusBukGraLupTou-JCP-16}, we can include the disregarded long-range MP2 correlation term in the KS-like equations by using the OEP scheme. In this manner, we define a fully self-consistent range-separated second-order OEP method (RS-OEP2) in which the orbitals are evaluated by the equations
\begin{eqnarray}
\left( -\frac{1}{2} \bm{\nabla}^2 +v_{\text{ne}}(\b{r}) + v_\text{H}(\b{r}) +  v_{\text{xc}}^{\text{RS-OEP2}}({\bf x})\right) \varphi_p({\bf x}) 
\nonumber\\
= \varepsilon_p \varphi_p({\bf x}),
\label{OEPRSHKS}
\end{eqnarray}
where the local exchange-correlation potential $v_{\text{xc}}^{\text{RS-OEP2}}$ is obtained by taking the functional derivative with respect to the density of all terms in Eq.~(\ref{ExcRSH})
\begin{eqnarray}\label{vxcOEPRSH}
v_{\text{xc}}^{\text{RS-OEP2}}({\bf x}) &=&   v_\text{xc}^\text{{sr,DFA}}({\bf x})  + v^\text{lr,EXX}_\text{x}({\bf x}) +  v_{\text{c}}^{\text{lr,GL2}}({\bf x}) \; .
\end{eqnarray}
The last two terms, namely $v_{\text{x}}^{\text{lr,EXX}}({\bf x}) = \delta E_\text{x}^\text{lr,HF}/\delta \rho({\bf x})$ and  $v_{\text{c}}^{\text{lr,GL2}}({\bf x}) = \delta E_\text{c}^\text{lr,MP2}/\delta \rho({\bf x})$ are the long-range exact exchange (EXX) and second-order G\"orling-Levy
(GL2) potentials, respectively. Note that GL2 is defined here without the single-excitation term, which is usually two orders of magnitude smaller than the double-excitation term~\cite{grabowski:2007:ccpt2,grabowski:2008:ijqc,grabowski:2014:jcp}. Since $E_\text{x}^\text{lr,HF}$ and $E_\text{c}^\text{lr,MP2}$ are only implicit functionals of the density through the orbitals and orbital energies, the calculation of $v_{\text{x}}^{\text{lr,EXX}}({\bf x})$ and $v_{\text{c}}^{\text{lr,GL2}}({\bf x})$ must be done with the OEP method~\cite{ivanov:1999:OEP,gorling:1999:OEP,grabowski:2014:jcp}. 
Thus, the long-range OEP exchange-correlation potential $v_{\text{xc}}^{\text{lr,OEP}}({\bf x}) = v_{\text{x}}^{\text{lr,EXX}}({\bf x}) + v_{\text{c}}^{\text{lr,GL2}}({\bf x})$ is given by the solution of the integral equation
\begin{equation}\label{e13}
\int v_{\text{xc}}^{\text{lr,OEP}}({\bf x}') \chi_\text{s}(\b{x}',\b{x}) \text{d}\b{x}' = \Lambda_\text{xc}^\text{lr}(\b{x})\ ,
\end{equation}
where $\chi_\text{s}(\b{x},\b{x}')$ is the static KS linear-response function and
\begin{eqnarray}
\Lambda_\text{xc}^\text{lr}(\b{x}) & = & \sum_p\Biggl[\int \text{d}\b{x}' \left( \funcder{E_\text{xc}^\lr}{\varphi_p(\b{x}')}\sum_{q\neq p}\frac{\varphi_q^*(\b{x})\varphi_p(\b{x})}{\varepsilon_p-\varepsilon_q}\varphi_q(\b{x}') + \text{c.c.}\right) 
\nonumber\\
&&+\frac{\partial E_\text{xc}^\lr}{\partial \varepsilon_p} |\varphi_p(\b{x})|^2\Biggl]\ ,
\end{eqnarray}
with $E_\text{xc}^\lr = E_\text{x}^\text{lr,HF} +  E_\text{c}^\text{lr,MP2}$. This equation is solved at each iteration of the self-consistent calculation of the orbitals in Eq.~(\ref{OEPRSHKS}). The obtained orbitals and orbital energies are then use to evaluate the RS-OEP2 energy using Eq.~(\ref{ERSH}).

\begin{figure}
\includegraphics[width=0.7\columnwidth]{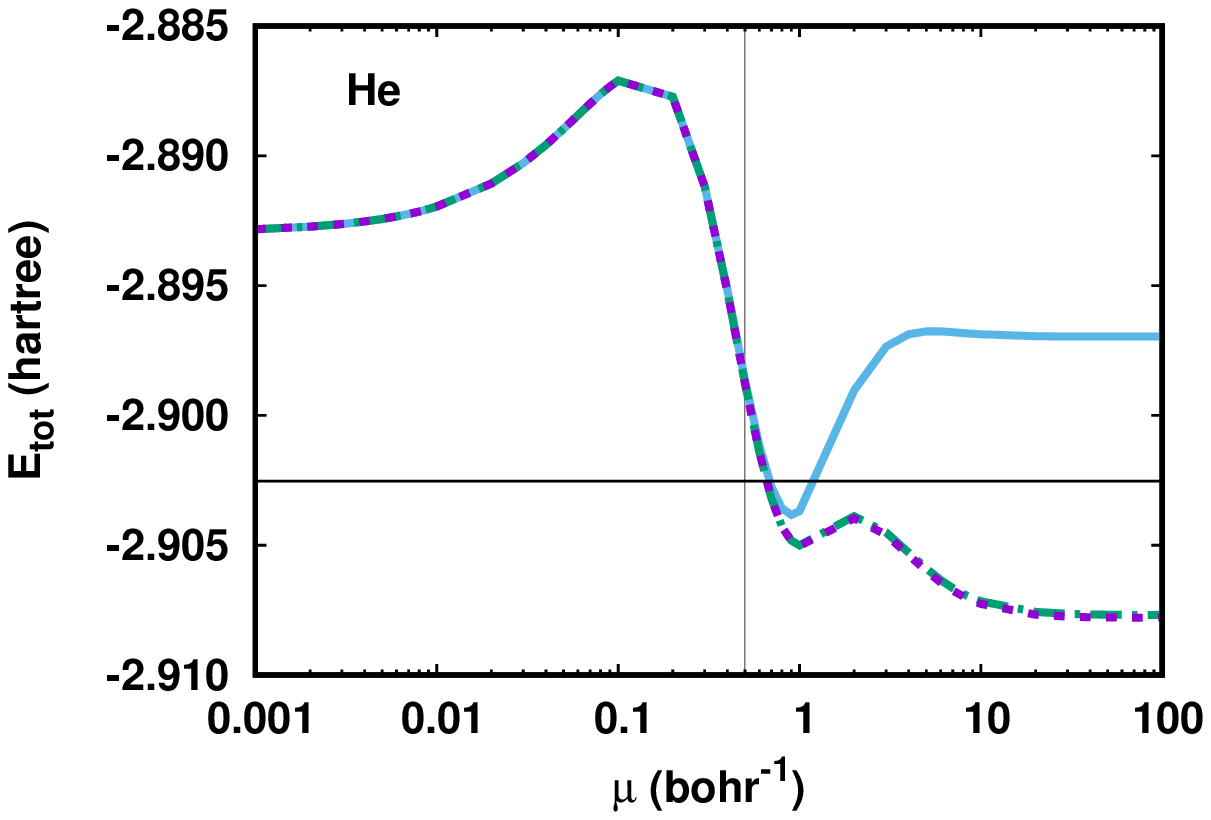}
\includegraphics[width=0.7\columnwidth]{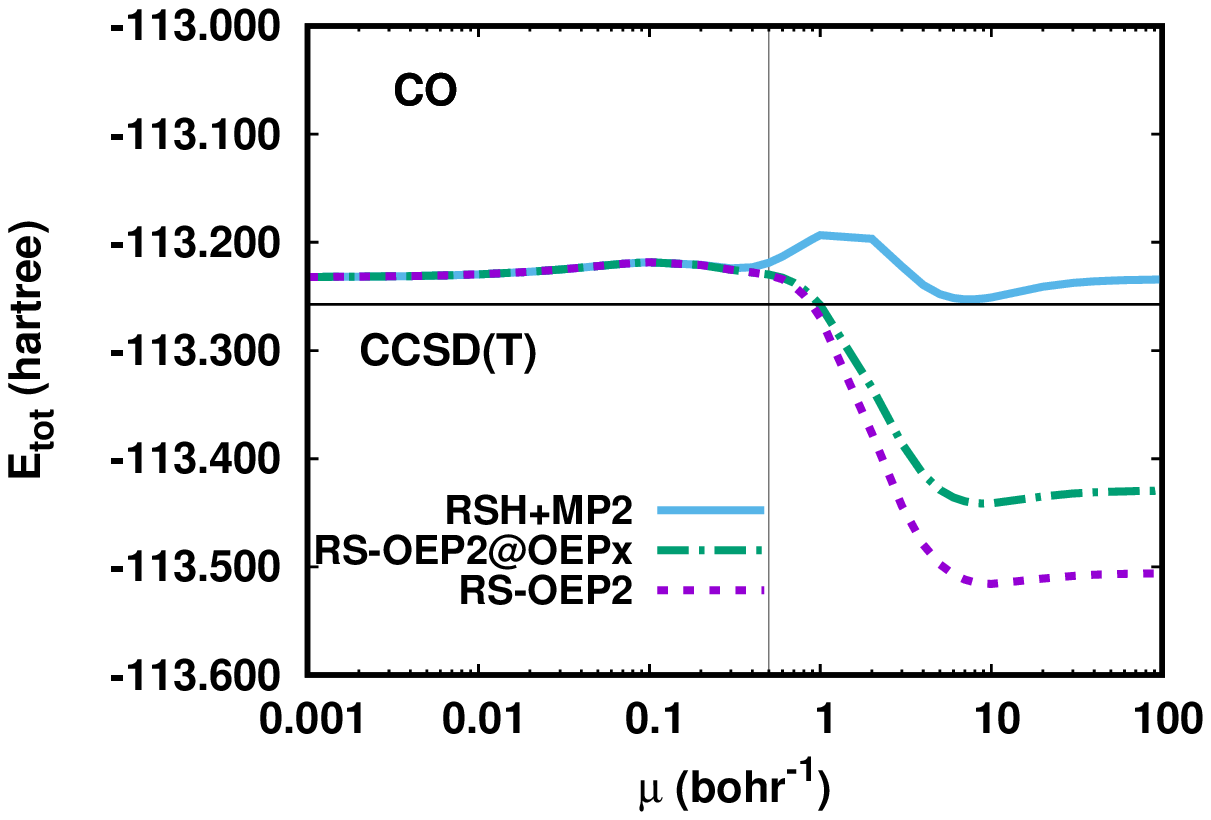}
\caption{
Total energies calculated with the RSH+MP2, RS-OEP2, and RS-OEP2@OEPx (where $v_{\text{c}}^{\text{lr,GL2}}({\bf x})$ was neglected when solving \Eq{OEPRSHKS}) methods with the srPBE exchange-correlation density functional as a function of the range-separation parameter $\mu$. The reference total energies are calculated with the CCSD(T) method in the same basis set (horizontal black line). The vertical lines correspond to commonly used value $\mu=0.5$ bohr$^{-1}$. 
}
\label{fig:etot2}
\end{figure}

\begin{figure}
	\includegraphics[scale=0.5]{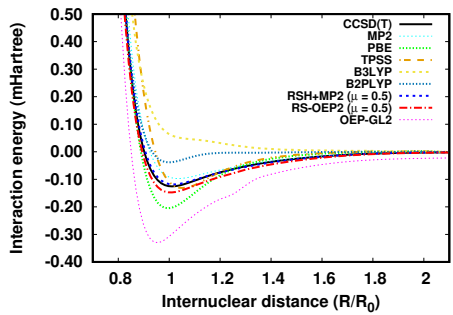}
	\caption{Interaction energy curves of the Ne dimer calculated using several approximations including RSH+MP2 and RS-OEP2 methods in the uncontracted aug-cc-pVTZ basis set. The reference results are calculated with the CCSD(T) method in the same basis set. 
	}
	\label{fig:ne2}
\end{figure}

\section{Computational details}
\label{sec:comput}

We have performed RSH+MP2 calculations with a development version of {\tt MOLPRO 2019}~\cite{Molproshort-PROG-19} and RS-OEP2 calculations with a development version of {\tt ACES II}~\cite{acesII} on a series of atoms and molecules (He, Be, Ne, Ar, CO, and H$_2$O). 

To calculate the long-range exchange-correlation OEP part of the potential, we employ the
finite basis-set procedure of \Refs{gorling:1999:OEP,ivanov:1999:OEP,ivanov:2002:OEP}.
Hence, in this approach, the OEP potential $v_{\text{xc}}^{\text{lr,OEP}}(\R)=v_{\text{xc}}^{\text{lr,OEP}}(\R\sigma)$, for spin-unpolarized systems, is expanded as
\begin{equation}\label{ee18}
v_{\text{xc}}^{\text{lr,OEP}}(\R) = v_{\text{Slater}}^{\text{lr}}(\R) + \sum_{l=1}^{M}c_l g_l(\R),
\end{equation}
where the first term is the long-range Slater potential, written in terms of spatial orbitals $\{\phi_{i}(\R)\}$,
\begin{equation}
v_{\text{Slater}}^{\text{lr}}(\R) = - \sum_{i,j}\frac{\phi_{i}^*(\R)\phi_{j}(\R)}{\rho(\R)}\int\phi_{i}(\R')\phi_{j}^*(\R') w_\text{ee}^\lr(|\R-\R'|) \text{d}\R',
\end{equation}
imposing the correct $-1/r$ asymptotic decay of the potential, and the second term is a correction spanned by $M$ auxiliary Gaussian basis functions $\{g_l(\R)\}$ with the expansion coefficients $\{c_l\}$ determined from the solution of the OEP equations. In order to compute the long-range Slater potential we have employ the resolution of the identity (RI) method proposed in \Ref{Sala-LHF-2001} combined with the modified interaction two-electron integrals. A similar approach was also used in \Ref{Krukau-LRSL-2008} to calculate the HF exchange energy density and the Slater potential. 

For the molecules, we considered the following equilibrium geometries: for
 CO $d(\text{C--O})=1.128 \text{\AA}$, 
and for  H$_2$O $d(\text{H--O})=0.959 \text{\AA}$ and $a(\text{H--O--H}) = 103.9^{\circ}$. In the OEP calculations, 
to ensure that the basis sets chosen were flexible enough for representation of orbitals and exchange-correlation 
potentials, all basis sets were constructed by full uncontraction of basis sets originally developed for 
correlated calculations, as in Refs.~\citenum{GraTeaSmiBar-JCP-11,Gra-MP-2014}. In particular, 
we employed an even tempered 20s10p2d basis for He, and an uncontracted ROOS--ATZP basis~\cite{widm90} for Be and Ne. 
For Ar, we used a modified basis set \cite{grabowski13} which combines s and p basis functions from the 
uncontracted ROOS--ATZP~\cite{widm90} with d and f functions coming from the uncontracted aug--cc--pwCVQZ basis 
set~\cite{peterson02}. In the case of both molecular systems, the uncontracted cc--pVTZ basis set
 of Dunning~\cite{dunning:1989:bas} was employed. Core excitations were included in the second-order correlation term.
 For the short-range DFA we have employed the short-range Perdew-Burke-Ernzerhof (srPBE) exchange-correlation density functional from \Ref{GolWerStoLeiGorSav-CP-06}.

For all OEP calculations standard convergence criteria were enforced,
 corresponding to maximum deviations in density-matrix elements of 10$^{-8}$. Additionally, the TSVD cutoff was set 
to 10$^{-6}$ and results were carefully checked to ensure convergence with respect to this parameter.
As in \Ref{grabowski:2014:jcp}, in order to check the stability of our solutions we have computed the gradient of the total electronic energy with respect to variations of the OEP expansion coefficients in \Eq{ee18}. For each value of $\mu$ the computed gradient had a norm less than 10$^{-12}$. Furthermore, the energies computed at slightly perturbed coefficients were higher than the one obtained in our converged calculations, which is consistent with an energy minimum.
 In the case of Be atom, which is strongly correlated, the OEP-GL2 calculation is unstable, as shown in Sec.~\ref{sec:results}. For more details regarding the OEP calculations, we refer to \Refs{SmiFraMusBukGraLupTou-JCP-16,grabowski:2014:jcp,SCSIP,BUK-2016}.
.
To assess the quality of the results obtained with the RSH+MP2 and RS-OEP2 methods, we have considered reference data from coupled-cluster singles doubles and perturbative triples [CCSD(T)]~\cite{purvis82,pople87,scuseria88,Raghavachari1989479} calculated in the same basis sets.

\begin{figure*}
\includegraphics[scale=0.45]{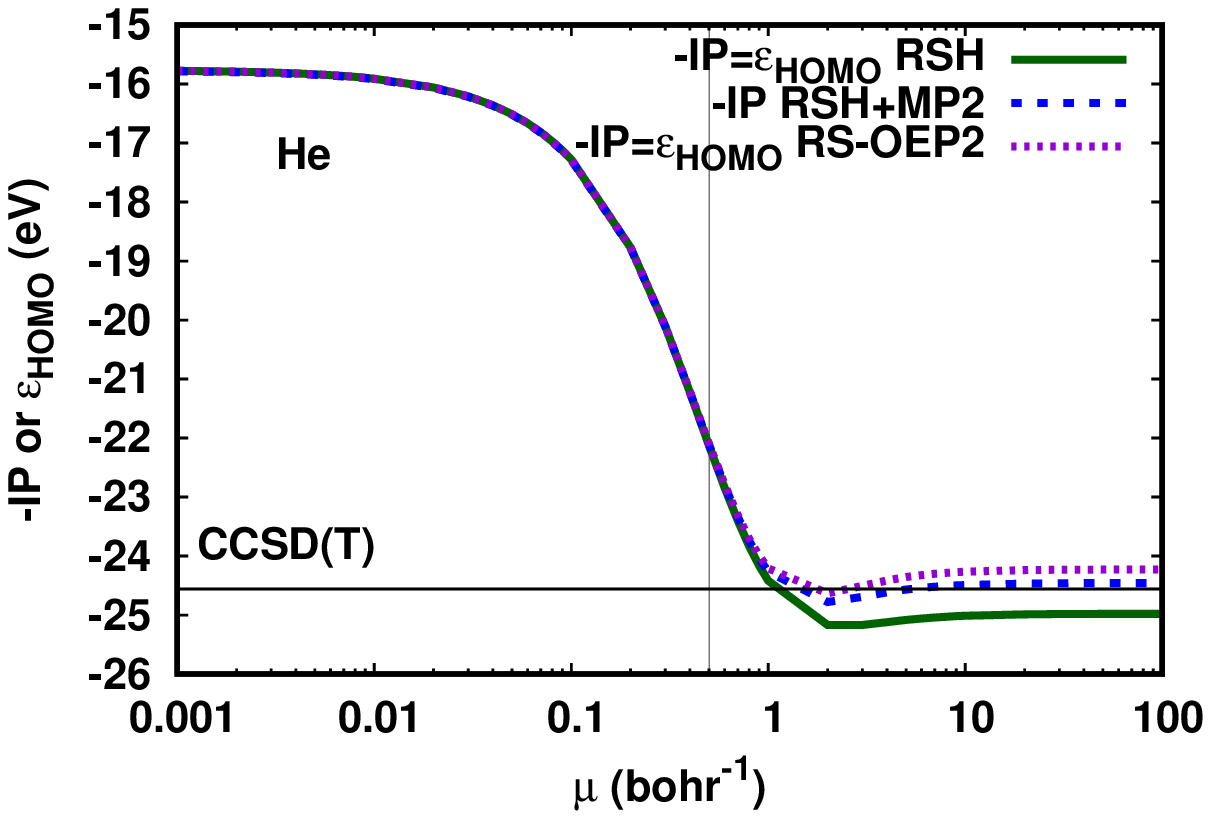}
\includegraphics[scale=0.45]{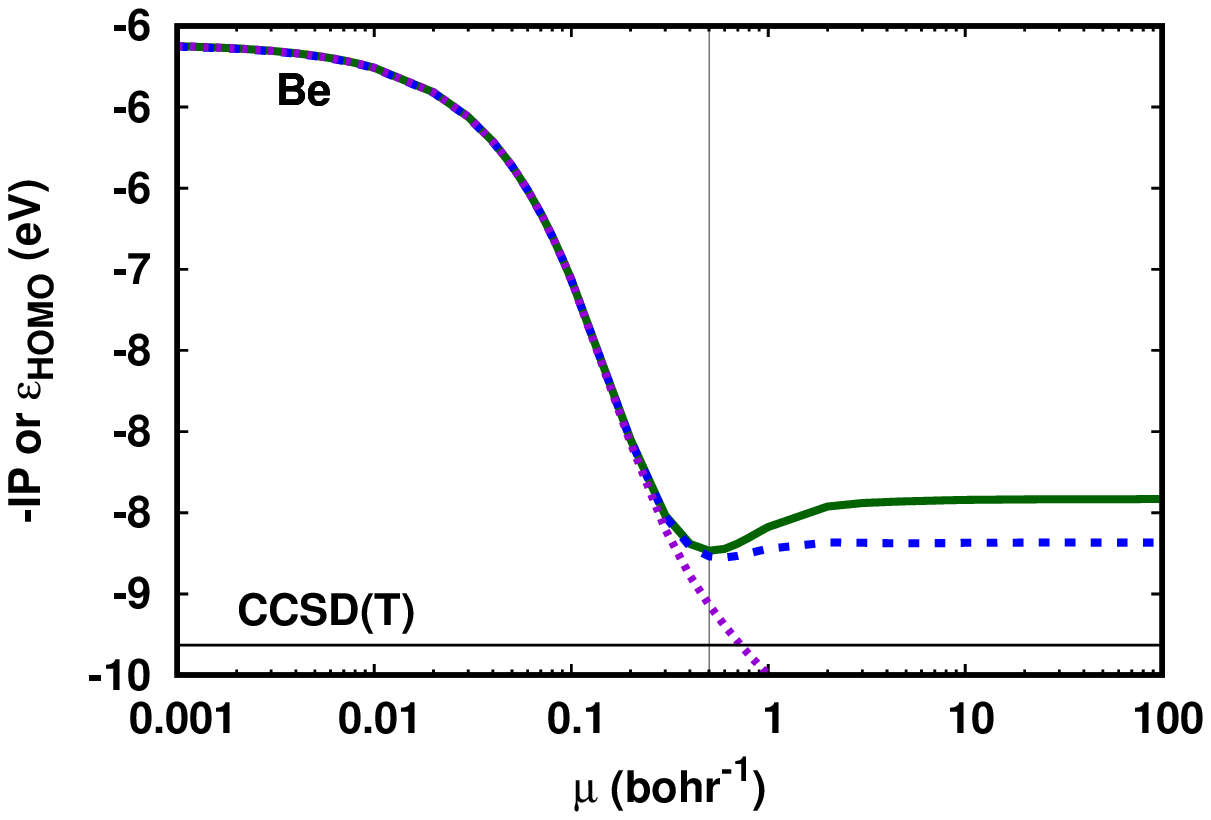}
\includegraphics[scale=0.45]{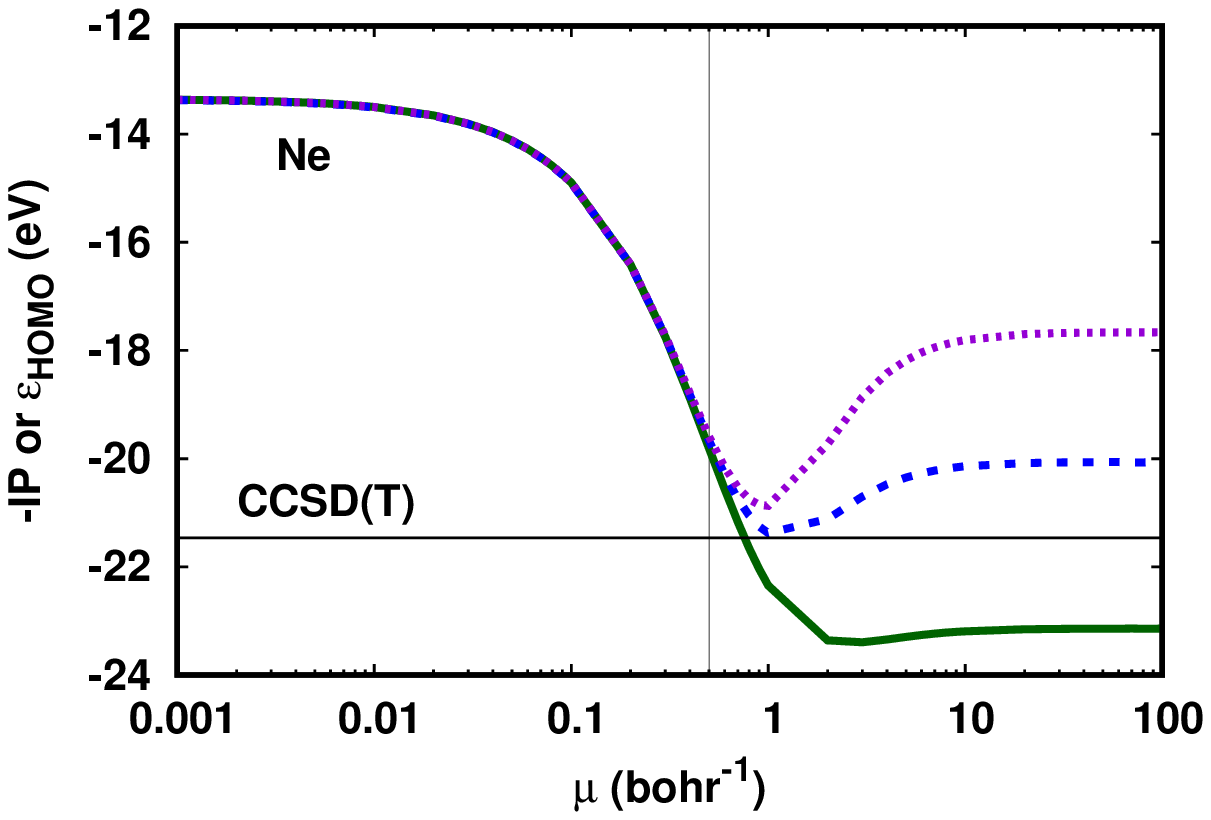}
\includegraphics[scale=0.45]{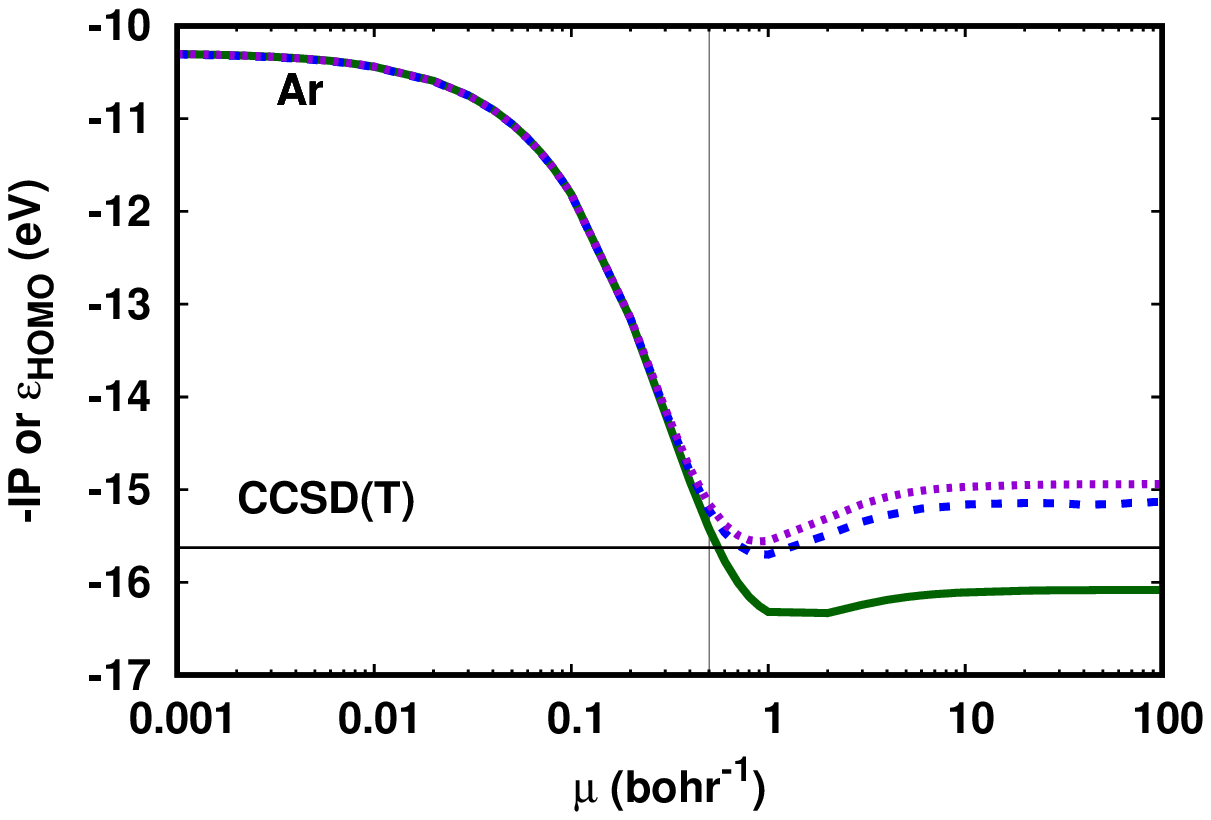}
\includegraphics[scale=0.45]{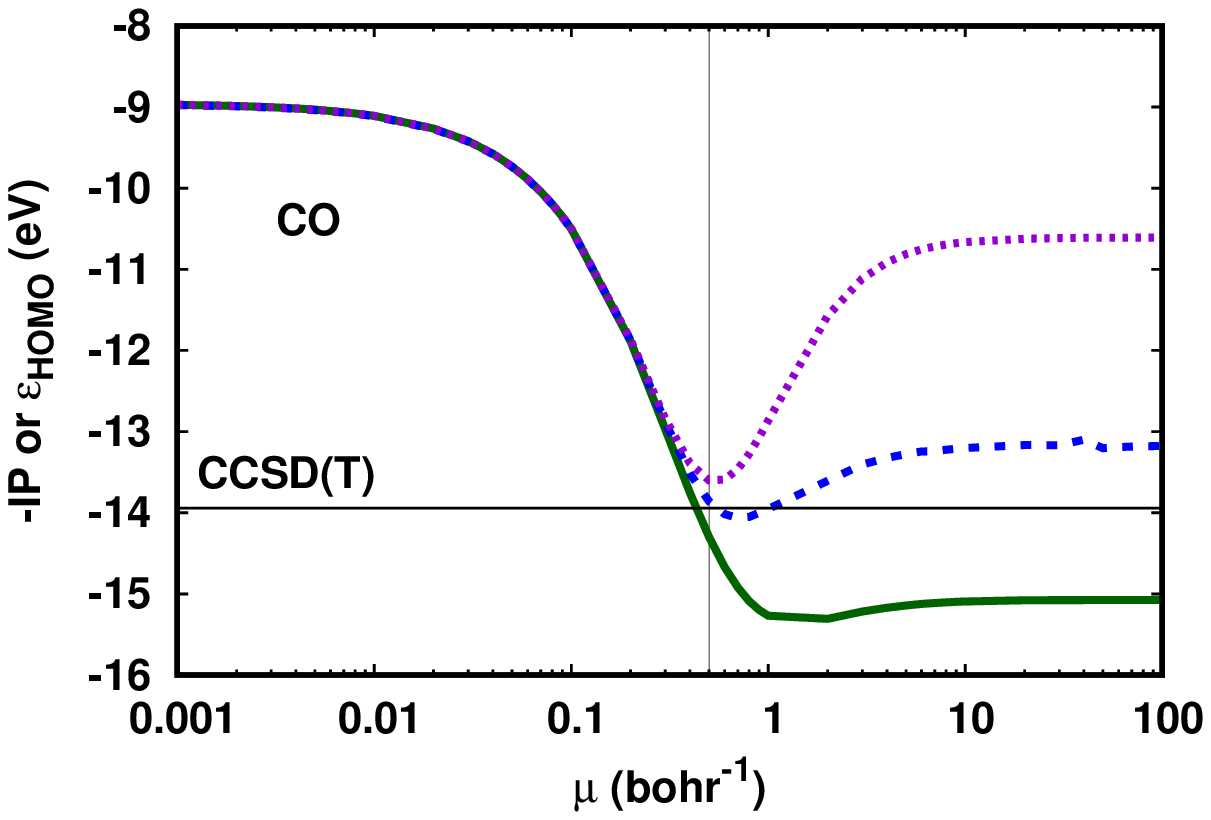}
\includegraphics[scale=0.45]{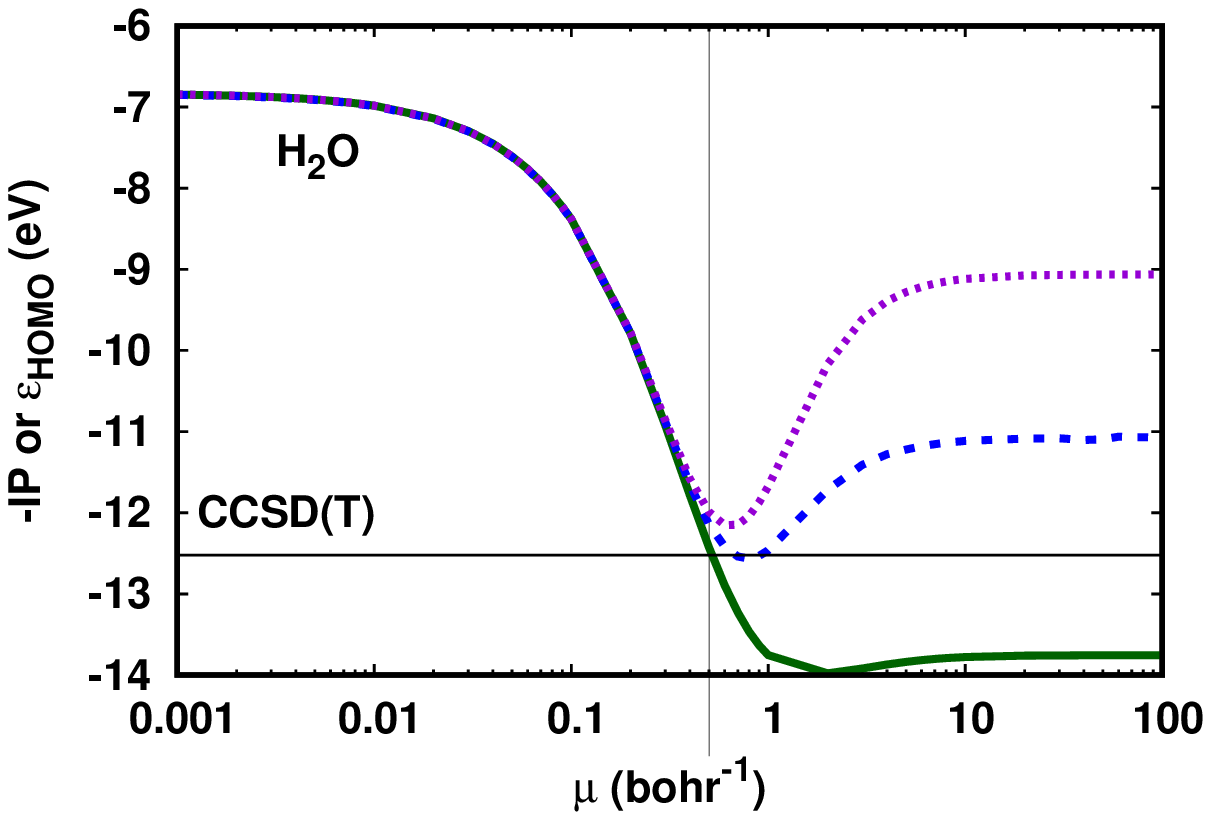}
\caption{Opposite of the IPs calculated by the RSH, RSH+MP2, and RS-OEP2 methods using the srPBE exchange-correlation functional as a function of range-separation parameter $\mu$. The reference values were calculated as CCSD(T) total energy differences with the same basis sets. The vertical lines correspond to commonly used value $\mu=0.5$ bohr$^{-1}$. For Be, the RS-OEP2 calculations are unstable for $\mu > 1$ bohr$^{-1}$.}
\label{fig:ip}
\end{figure*}
\begin{table*}[htbp]
\caption{IPs (in eV) calculated with the RSH+MP2 and RS-OEP2 methods using the srPBE exchange-correlation density functional for different range-separation parameters $\mu$ (in bohr$^{-1}$). The reference CCSD(T) values were calculated as total energy differences with the same basis sets. The last line gives the mean absolute error (MAE) with respect to the CCSD(T) values.}
\begin{tabular}{lcccccccccc}
\hline \hline
& CCSD(T) & \multicolumn{ 4}{c}{RSH+MP2} & & \multicolumn{ 4}{c}{RS-OEP2}   \\
\cline{3-6} \cline{8-11}
       &        & $\mu =  0$ &  $\mu = 0.5$ & $\mu = 1$ &  $\mu \to \infty$ &  & $\mu =  0$ &  $\mu = 0.5$ & $\mu = 1$ &  $\mu \to \infty$ \\ \hline
He     &  24.56  & 15.79 &  22.11 & 24.27 & 24.46 &&   15.79 &  22.10 &  24.21 &  24.23   \\ 
Be     &  9.31   & 5.63  &  8.77  & 8.72  & 8.68  &&   5.63  &  9.07  &  9.50  & -        \\ 
Ne     &  21.47  & 13.37 &  19.66 & 21.37 & 20.07 &&   13.37 &  19.59 &  20.87 &  17.66   \\
Ar     &  15.63  & 10.31 &  15.22 & 15.70 & 15.14 &&   10.31 &  15.16 &  15.55 &  14.94   \\ 
CO     &  13.94  & 8.98  &  13.86 & 13.95 & 13.17 &&   8.98  &  13.60 &  12.86 &  10.61   \\ 
H$_2$O &  12.52  &  6.85 &  12.14 & 12.46 & 11.07 &&   6.85  &  11.99 &  11.67 &  9.06    \\[0.1cm]
MAE    &         & 6.09  &   0.94 & 0.19  & 0.81  &&  6.09   & 0.99   & 0.52   & 2.32     \\ \hline \hline
\end{tabular}
\label{tab:ip}
\end{table*}

\section{Results and discussion}
\label{sec:results}

\subsection{Total energies}

In \Fig{fig:etot} we report the total energies of each system as a function of the range-separation parameter $\mu$ calculated using the RSH+MP2 and RS-OEP2 methods. At $\mu=0$ both methods reduce to standard KS DFT with the PBE exchange-correlation functional. In this limit, the total energies are systematically too high (up to about 100 mHa) in comparison with the CCSD(T) reference values. When $\mu \rightarrow \infty$, RSH+MP2 reduces to standard MP2, and RS-OEP2 reduces to a full-range OEP-GL2 calculation. For all systems, standard MP2 gives total energies which are too high by a modest amount (at most about 20 mHa). By contrast, OEP-GL2 gives much too negative total energies (by more than 200 mHa for CO) which is a consequence of the large overestimation of the correlation effects by this approximation~\cite{Grab-JCP-2011,Gra-MP-2014, grabowski:2014:jcp}. 
This feature has two origins: i) the much smaller HOMO-LUMO gap obtained from the KS calculations where both the occupied and unoccupied orbital energies are defined with the same local effective potential~\cite{MorWuYan-JCP-05,Eng-OEP2-2005}; ii) the optimization of orbitals conducted in the presence of the correlation potential $v_{\text{c}}^{\text{lr,GL2}}({\bf x})$ which leads to a further amplification of the overestimation due to its explicit dependence on the KS orbital energies. This can be seen in \Fig{fig:etot2} where we compare, for two representative systems (He and CO), the results obtained with RSH+MP2, RS-OEP2, and a simplified RS-OEP2 variant (denoted as RS-OEP2@OEPx) in which only the long-range EXX potential is calculated using the OEP procedure and the long-range MP2 correlation energy is calculated a posteriori like in the RSH+MP2 case. We note that RS-OEP2@OEPx already largely overestimates the reference CCSD(T) results. For He, RS-OEP2 and RS-OEP2@OEPx give almost the same total energies, much lower than the RSH+MP2 total energy for large $\mu$ due to the smaller HOMO-LUMO gap value. For CO, this effect is further increased by the orbital relaxation in the self-consistent procedure including the correlation potential $v_{\text{c}}^{\text{lr,GL2}}({\bf x})$. We note that, for intermediate values of $\mu$, the RSH+MP2 and RS-OEP2 total energies show a non-monotonic variation with respect to $\mu$. At the scale of the plots, the two methods give essentially identical total energies for $\mu \lesssim 0.2-1$ bohr$^{-1}$. The difference between the two methods becomes important for $\mu \gtrsim 1$ bohr$^{-1}$ where the long-range exchange-correlation contribution starts to dominate over the srPBE exchange-correlation functional. This fact actually explains the lack of convergence of RS-OEP2 for $\mu > 1$ bohr$^{-1}$ in the case of the Be atom. The dominating role of the GL2 part of the potential causes the HOMO-LUMO gap to close and leads to an instability for $\mu > 1$ bohr$^{-1}$. This aspect was investigated in \Refs{MorWuYan-JCP-05,Eng-OEP2-2005}.
For the commonly used value of the range-separation parameter, $\mu = 0.5$ bohr$^{-1}$,~\cite{GerAng-CPL-05a,MusReiAngTou-JCP-15} the RSH+MP2 and RS-OEP2 methods thus perform overall very similarly for total energies. 

{In \Fig{fig:ne2} we report the interaction energy curves for the Ne dimer calculated using the RSH+MP2 and RS-OEP2 methods. For comparison we also report the curves obtained with several approximations from several rungs of Perdew's ladder, namely GGA (PBE, $\mu=0$), meta-GGA (TPSS~\cite{TaoPerStaScu-PRL-03}), hybrid (B3LYP~\cite{Bec-JCP-93,BarAda-CPL-94,SteDevChaFri-JPC-94}) and double hybrid (B2PLYP~\cite{Gri-JCP-06}). Additionally, as a reference, we have added the CCSD(T) curve. As one can see, RSH+MP2 slightly underbinds and RS-OEP2 overbinds the Ne dimer, which is consistent with a larger overestimation of the total energy by RS-OEP2 in the dimer compared to the atom. Nevertheless both methods improve over their original counterparts, namely MP2 and OEP-GL2. We can also note the rather large improvement over the other types of approximations.

\begin{figure*}
\includegraphics[scale=0.45]{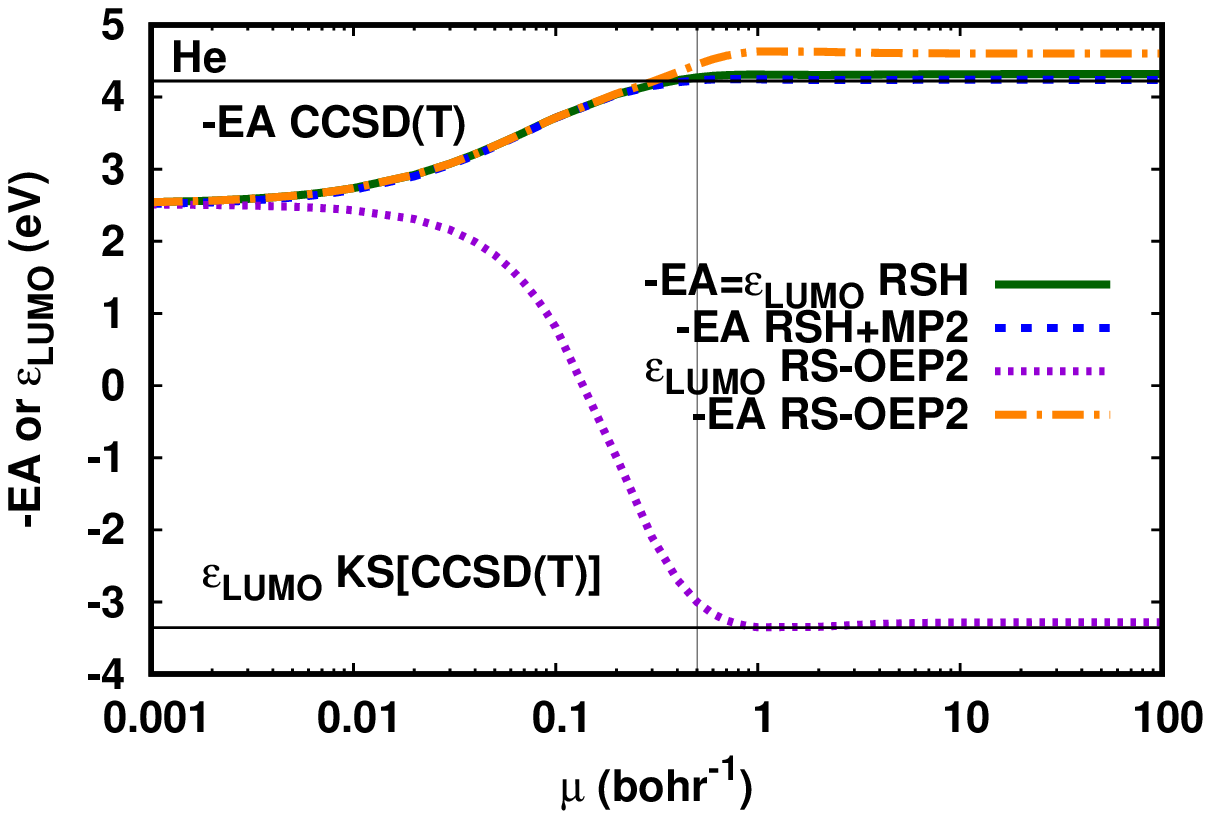}
\includegraphics[scale=0.45]{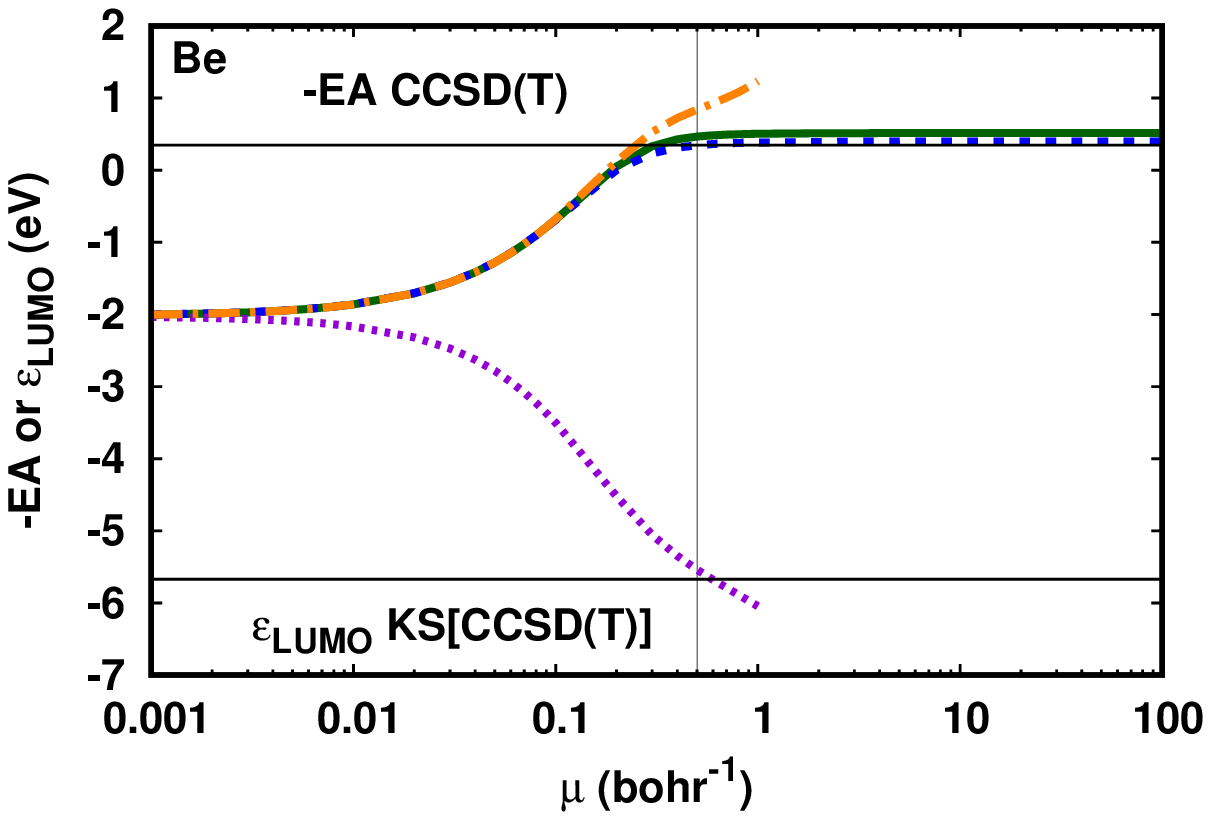}
\includegraphics[scale=0.45]{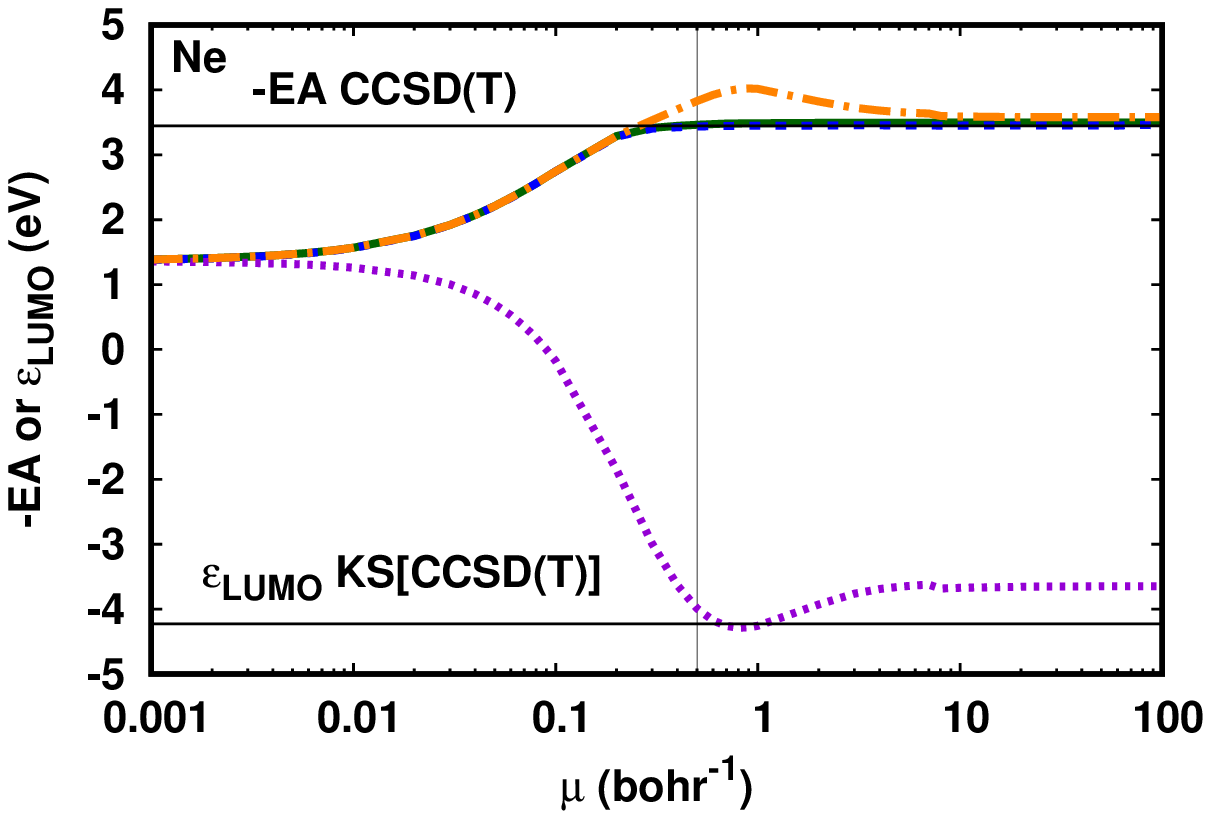}
\includegraphics[scale=0.45]{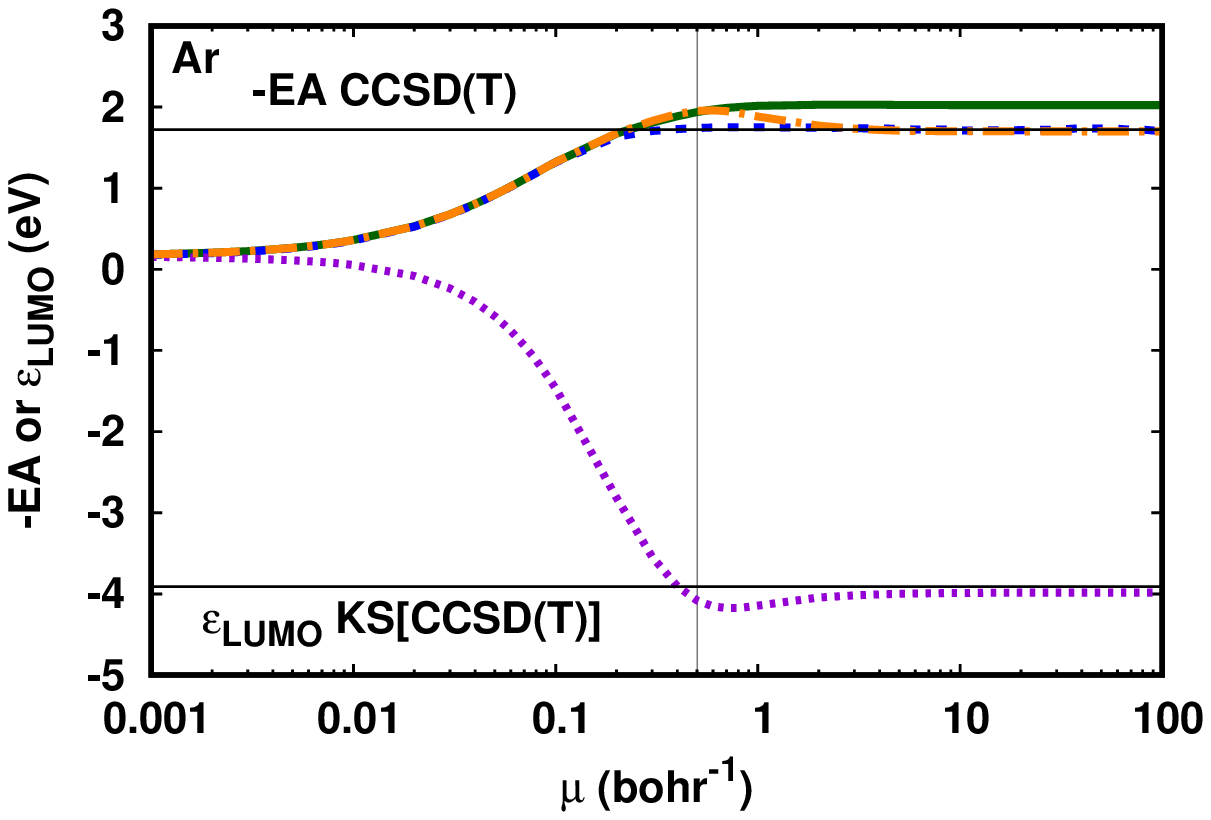}
\includegraphics[scale=0.45]{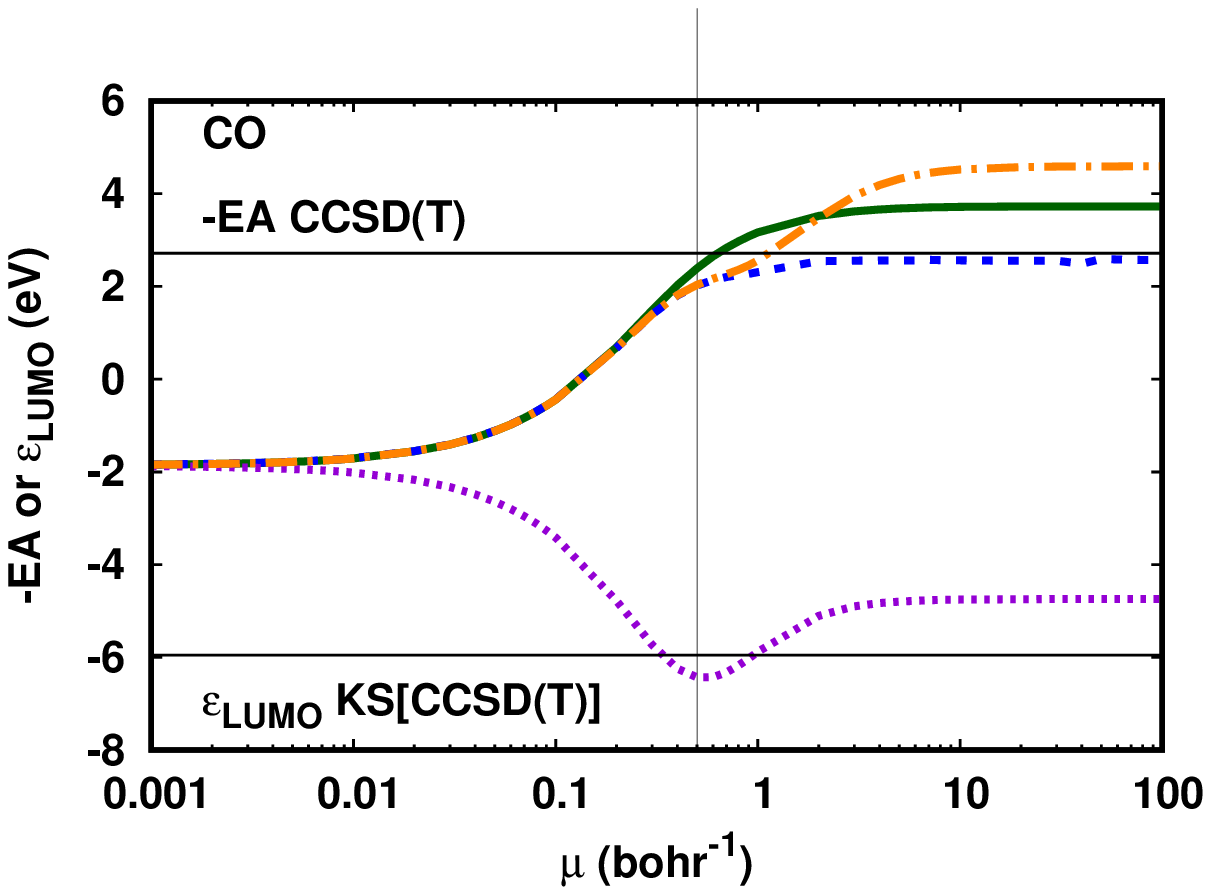}
\includegraphics[scale=0.45]{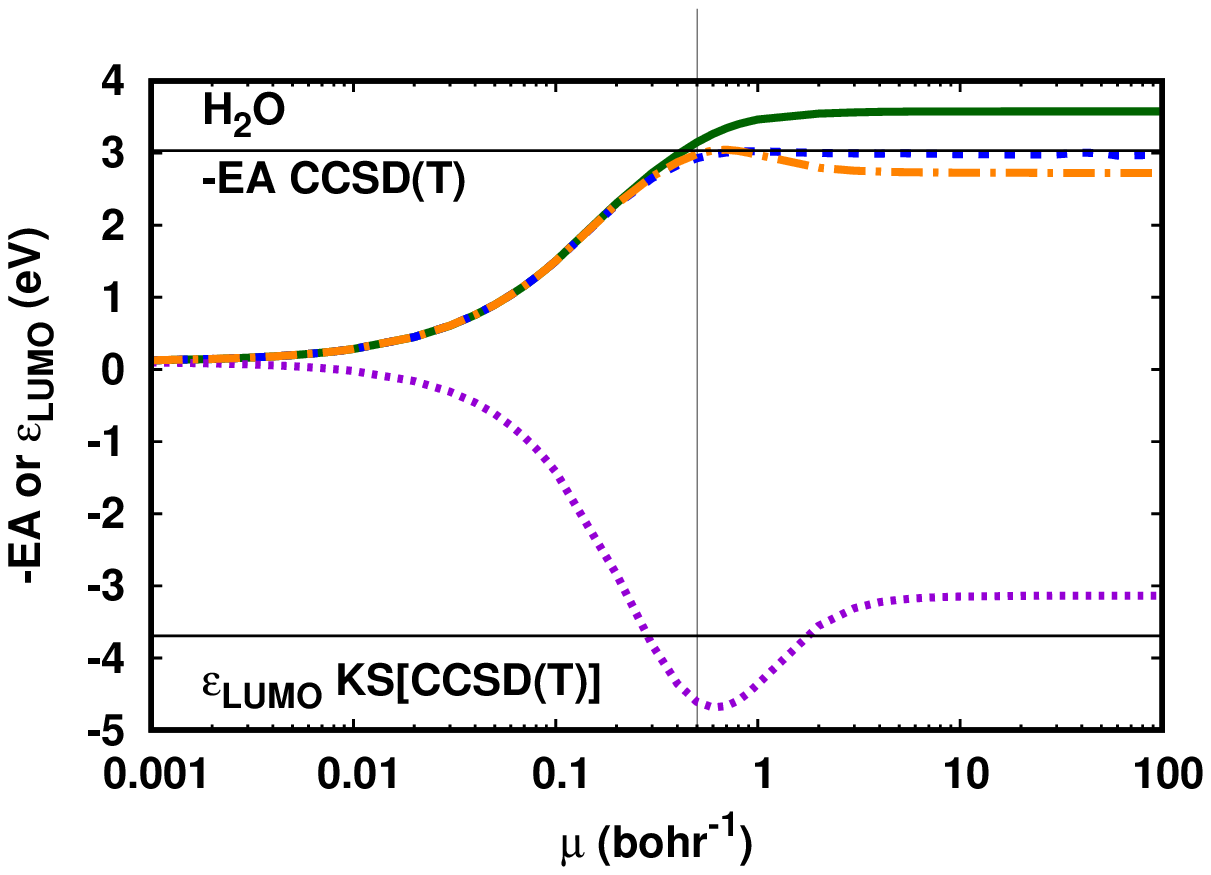}
\caption{Opposite of the EAs calculated by the RSH, RSH+MP2, and RS-OEP2 methods, together with LUMO orbital energies for the RS-OEP2 method, using the srPBE exchange-correlation functional as a function of range-separation parameter $\mu$. The reference values were calculated as CCSD(T) total energy differences with the same basis sets. The vertical lines correspond to commonly used value $\mu=0.5$ bohr$^{-1}$. For Be, the RS-OEP2 calculations are unstable for $\mu > 1$ bohr$^{-1}$.}
\label{fig:ea}
\end{figure*}
\begin{table*}[htbp]
\caption{Opposite of EAs (in eV) calculated with the RSH+MP2 and RS-OEP2 methods using the srPBE exchange-correlation density functional for different range-separation parameters $\mu$ (in bohr$^{-1}$). The reference CCSD(T) values were calculated as total energy differences with the same basis sets. The last line gives the mean absolute error (MAE) with respect to the CCSD(T) values.}
\begin{tabular}{lcccccccccc}
\hline \hline
& CCSD(T) & \multicolumn{ 4}{c}{RSH+MP2} & & \multicolumn{ 4}{c}{RS-OEP2}   \\
\cline{3-6} \cline{8-11}
      &      & $\mu =  0$   &  $\mu = 0.5$ & $\mu = 1$ &  $\mu \to \infty$ &  & $\mu =  0$ &  $\mu = 0.5$ & $\mu = 1$ &  $\mu \to \infty$ \\ \hline
He    & 4.22 &  2.52  & 4.23 & 4.25 & 4.24 &&2.52  & 4.45 & 4.63 & 4.60  \\
Be    & 0.35 &  -2.00 & 0.35 & 0.38 & 0.40 &&-2.00 & 0.84 & 1.23 & -     \\ 
Ne    & 3.44 &  1.38  & 3.44 & 3.45 & 3.46 &&1.38  & 3.83 & 4.01 & 3.58  \\ 
Ar    & 1.72 &  0.18  & 1.74 & 1.75 & 1.71 &&0.18  & 1.95 & 1.88 & 1.70  \\
CO    & 2.72 &  -1.84 & 2.02 & 2.31 & 2.57 &&-1.84 & 2.03 & 2.56 & 4.59  \\ 
H$_2$O& 3.04 &  0.13  & 2.93 & 3.02 & 2.98 &&0.13  & 2.99 & 2.97 & 2.73  \\[0.1cm] 
MAE   &      &  2.52  & 0.14 & 0.09 & 0.05 &&2.52  & 0.35 & 0.38 & 0.55  \\ \hline \hline
\end{tabular}
\label{tab:ea}
\end{table*}

\subsection{HOMO orbital energies and ionization potentials}
\label{sec:ip}
\begin{figure*}
\includegraphics[scale=0.45]{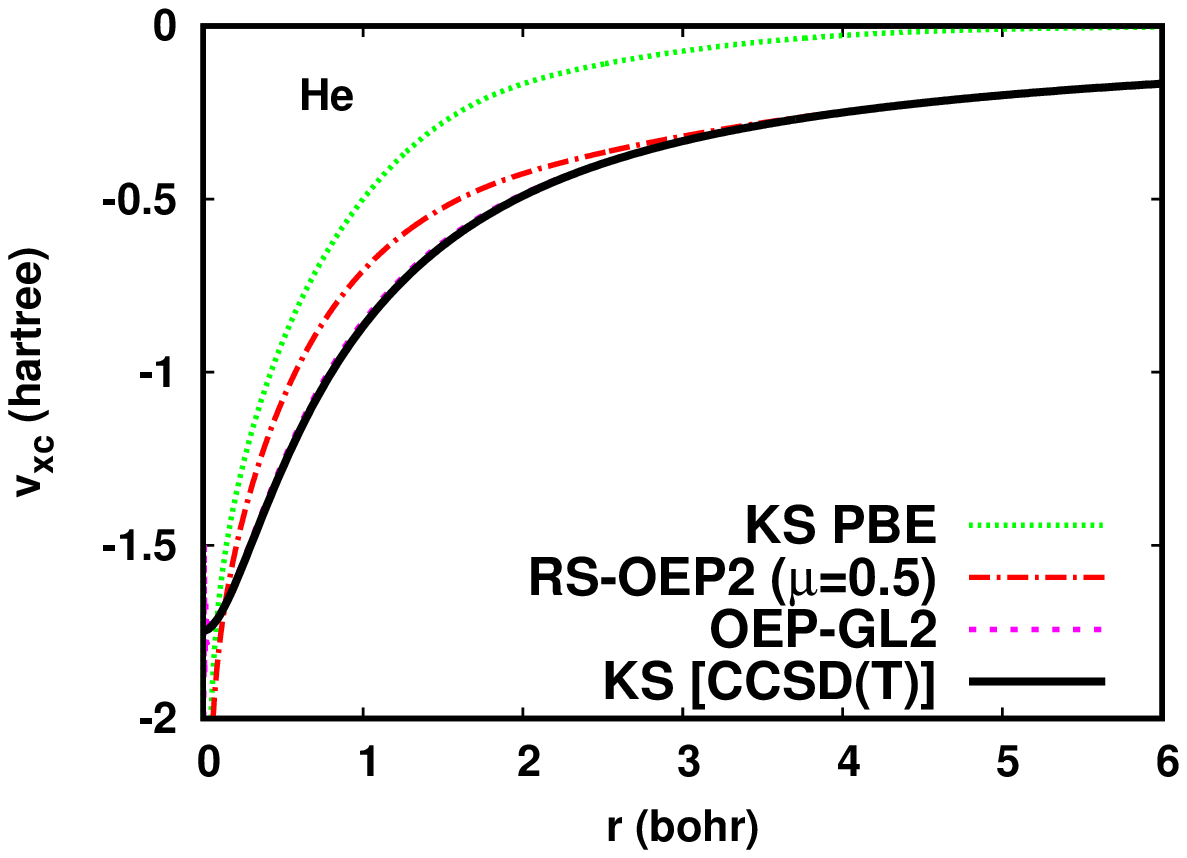}
\includegraphics[scale=0.45]{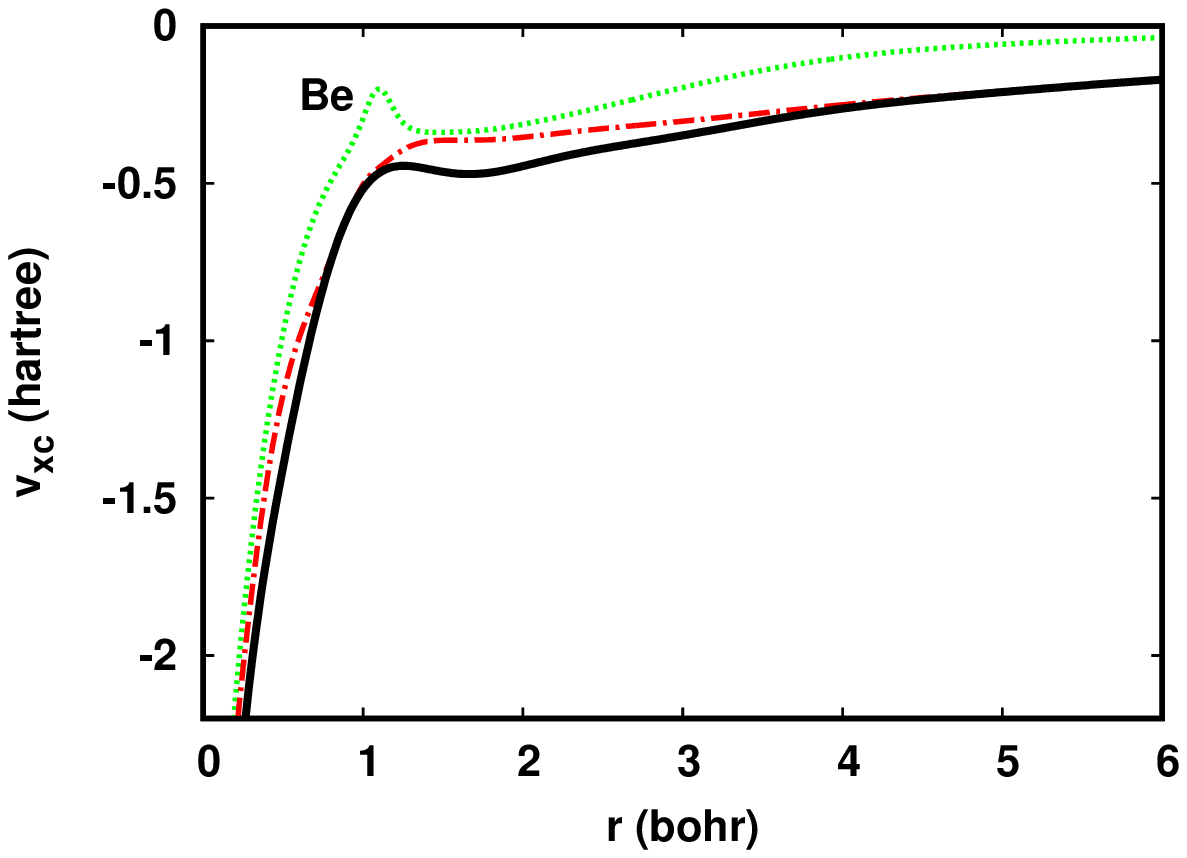}
\includegraphics[scale=0.45]{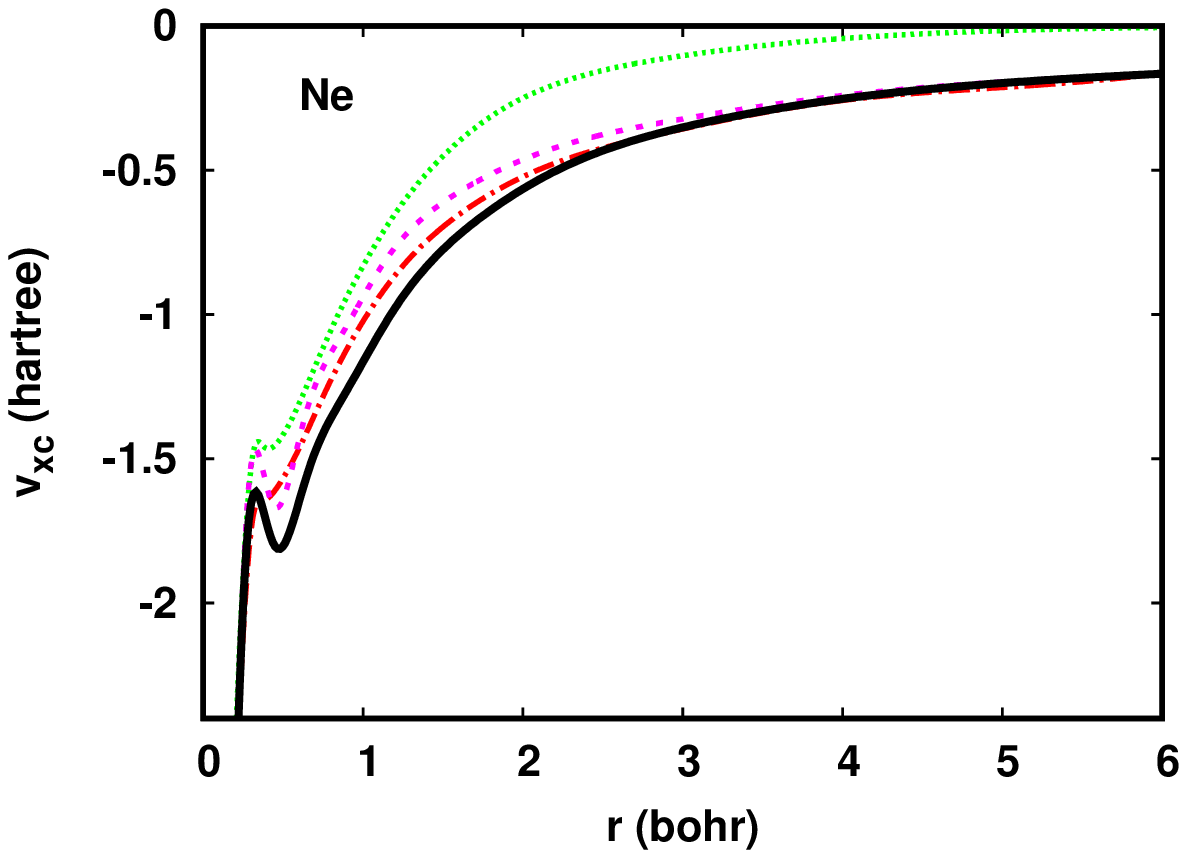}
\includegraphics[scale=0.45]{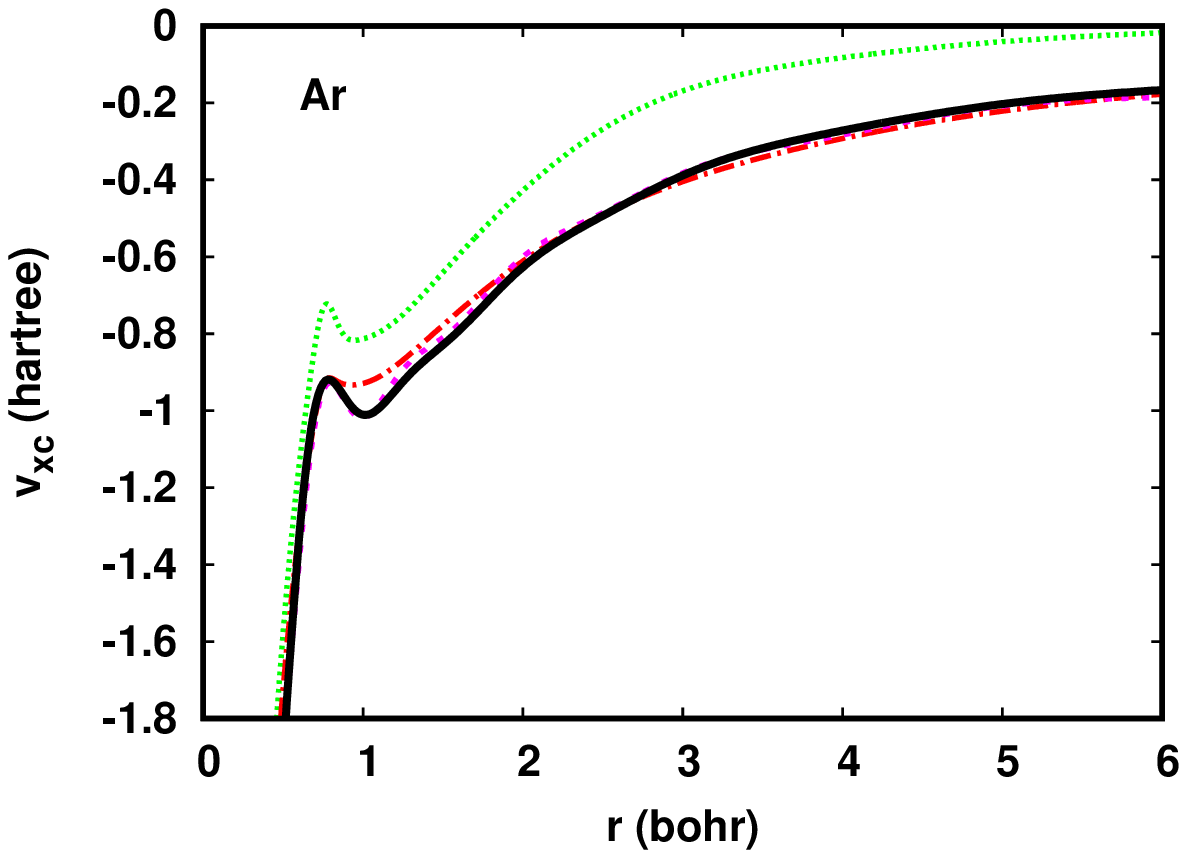}
\includegraphics[scale=0.45]{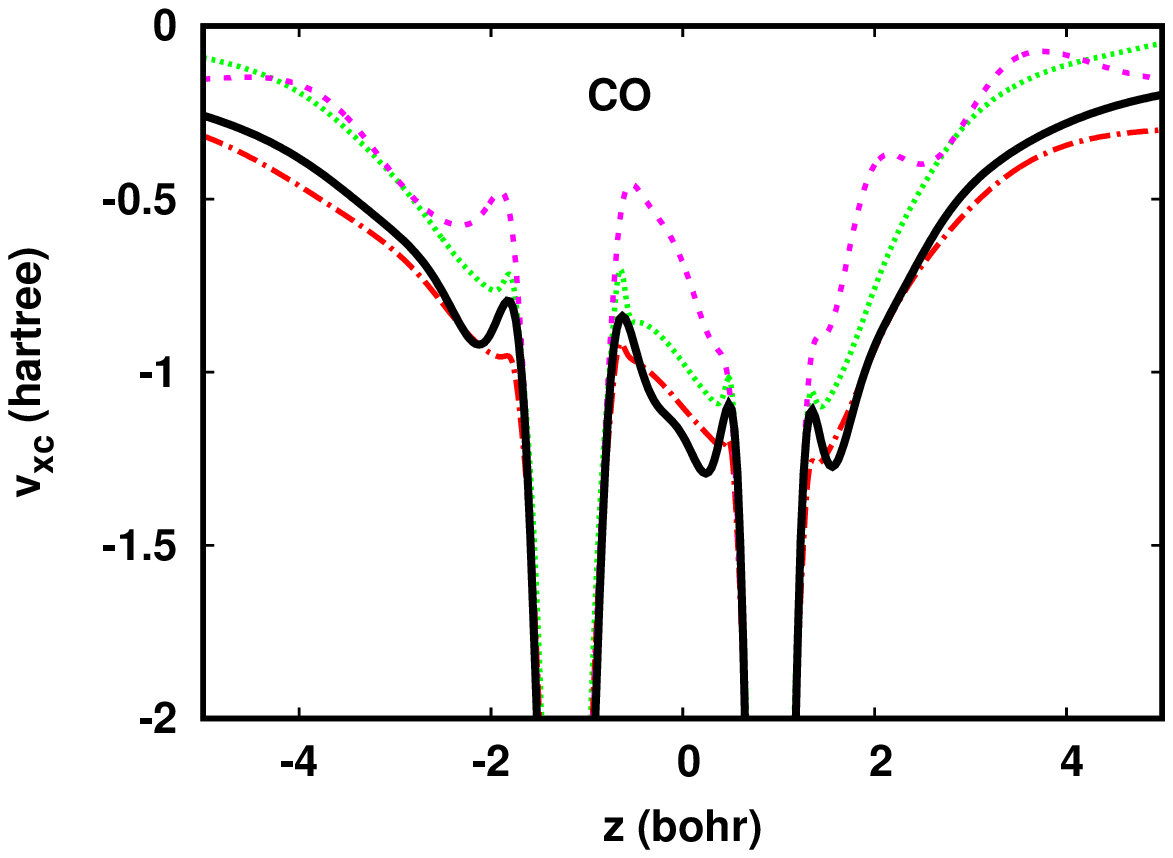}
\includegraphics[scale=0.45]{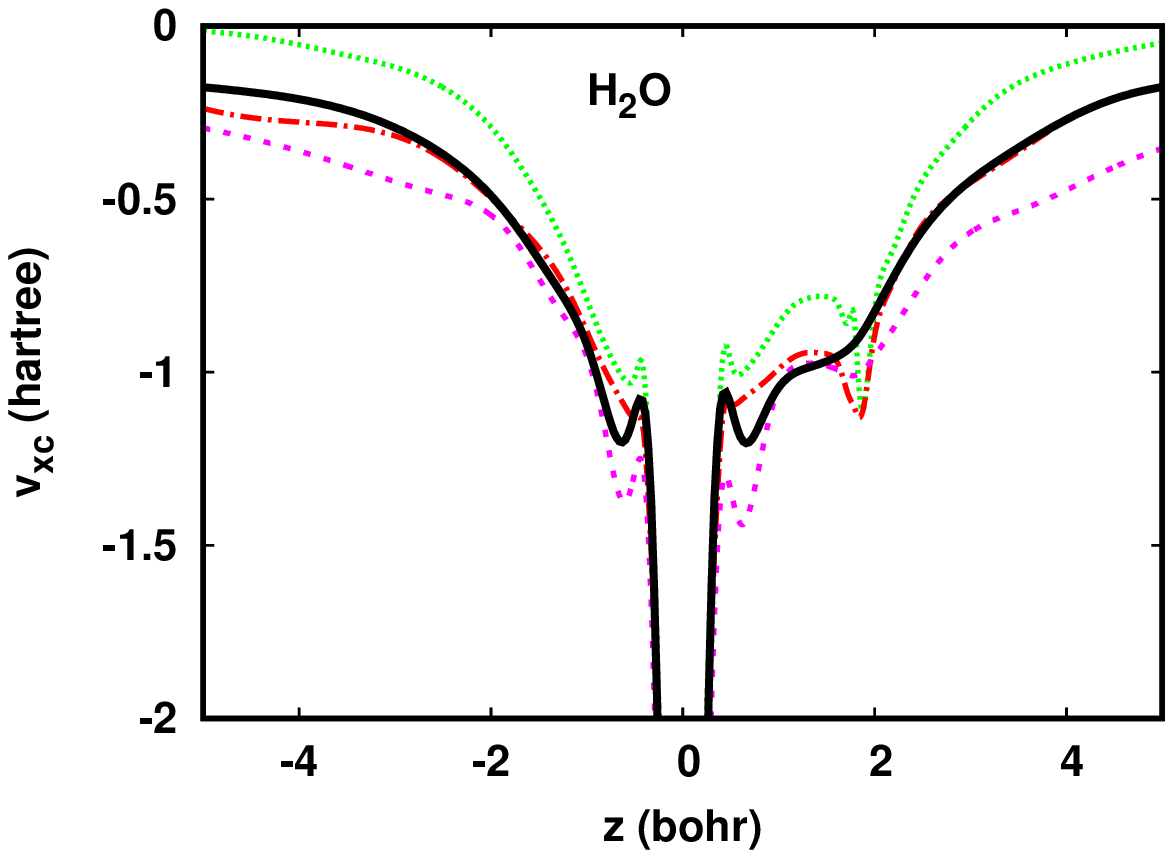}
\caption{Exchange-correlation potentials calculated with the RS-OEP2 method using the srPBE exchange-correlation density functional for the range-separation parameter $\mu=0.5$ bohr$^{-1}$, and for the limiting values $\mu = 0$ (standard KS PBE) and $\mu \to \infty$ (OEP-GL2). The reference potentials were calculated by KS inversion of CCSD(T) densities. For Be, the OEP-GL2 calculation is unstable.
For CO the correlation potentials are calculated along the molecular axis,
and for H$_2$O along the HO bond.}
\label{fig:vxc}
\end{figure*}

In \Fig{fig:ip}, we report the ionization potentials (IP) calculated with the RSH+MP2 and RS-OEP2 methods as a function of $\mu$. We also report the IPs obtained by the RSH method which simply corresponds to dropping the MP2 correlation term in the RSH+MP2 method and constitutes the first step of a RSH+MP2 calculation. For the non-self-consistent RSH+MP2 method, the IP is calculated as a finite-difference derivative of the total energy with respect to the electron number, as described in \Refs{CohMorYan-JCTC-09,SuYanMorXu-JPCA-14,SmiFraMusBukGraLupTou-JCP-16}. For the self-consistent RSH and RS-OEP2 methods, the IP is simply obtained as the opposite of the HOMO orbital energy~\cite{perdew82,levy84,amb85}. As in \Ref{SmiFraMusBukGraLupTou-JCP-16}, in all RS-OEP2 calculations, the HOMO condition on the OEP exchange \cite{kli,ivanov:2002:OEP, gorling:1999:OEP} and correlation \cite{grabowski:2002:OEP2,SCSIP} potentials has been imposed. The reference values (black horizontal line) were calculated using the CCSD(T) method as the difference between the total energies of the $N$- and $(N-1)$-electron systems using the same basis sets.

For $\mu = 0$, all methods reduce to KS PBE which gives a HOMO energy systematically higher than the reference CCSD(T) $-$IP value by about 4 to 9 eV (depending on the considered system). This must be due to self-interaction error introduced by this semilocal DFA. For $\mu \to \infty$, the RSH method reduces to HF with a HOMO energy which provides a much better estimation of $-$IP, differing from the reference value by about $1-2$ eV. In the same limit, RSH+MP2 reduces to standard MP2 which tends to give IPs even closer to the reference CCSD(T) values. In the case of the RS-OEP2 method, when $\mu\to \infty$, we obtain the OEP-GL2 method which tends to give substantially too high HOMO energies. 

For intermediate values of $\mu$, we first note the non-monotonic behavior of all the curves. Furthermore, in the range of $\mu$ between about $0.4$ and $2.0$ bohr$^{-1}$ (depending on the system), almost all the curves have minima which for RSH+MP2 and RS-OEP2 correspond to IPs relatively close to the reference CCSD(T) values. The only exception is the Be atom for which the RS-OEP2 HOMO and LUMO energies (see \Sec{sec:ea}) rapid decreases with $\mu$ just before the calculation becomes unstable for $\mu > 1$ bohr$^{-1}$.

In order to investigate more closely the performance of the RSH+MP2 and RS-OEP2 methods in the vicinity of the minimum, we report in \Tab{tab:ip} the IPs calculated for several values of the range-separation parameter $\mu$. For $\mu = 0$, where the two methods reduce to KS PBE, the mean absolute error (MAE) is as large as 6.09 eV. For the commonly used value $\mu = 0.5$ bohr$^{-1}$, RSH+MP2 and RS-OEP2 give similar MAEs of 0.94 and 0.99 eV, respectively. For $\mu = 1$ bohr$^{-1}$, we observe a further decrease of the MAEs, with 0.19 and 0.52 eV for RSH+MP2 and RS-OEP2, respectively. A similar behavior was also observed in \Ref{Hesselmann2018} for the self-consistent OEP version of a range-separated RPA method. The better performance of RSH+MP2 over RS-OEP2 for this larger value of $\mu$ is consistent with the $\mu\to\infty$ limit where standard MP2 gives more accurate IPs than OEP-GL2 (MAEs of 0.81 and 2.32 eV, respectively). 

We thus conclude that self-consistency does not bring any improvement for the calculation of IPs in the RSH+MP2 approach. More accurate IPs might be obtained by using a spin-component-scaled second-order correlation energy expression~\cite{SCSIP} both in RS-OEP2 and in RSH+MP2~\cite{Smi-JCTC-2018}.

\subsection{LUMO orbital energies and electronic affinities}
\label{sec:ea}

\begin{figure*}
\includegraphics[scale=0.45]{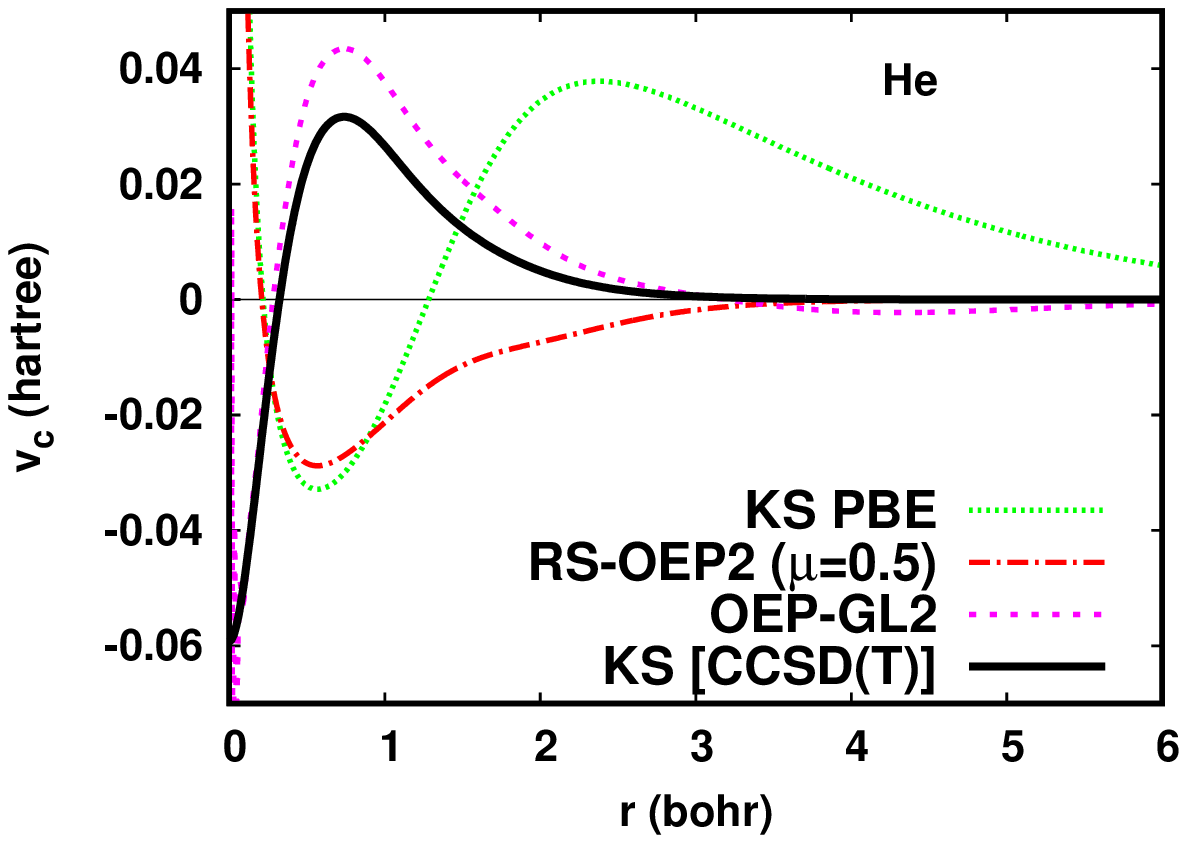}
\includegraphics[scale=0.45]{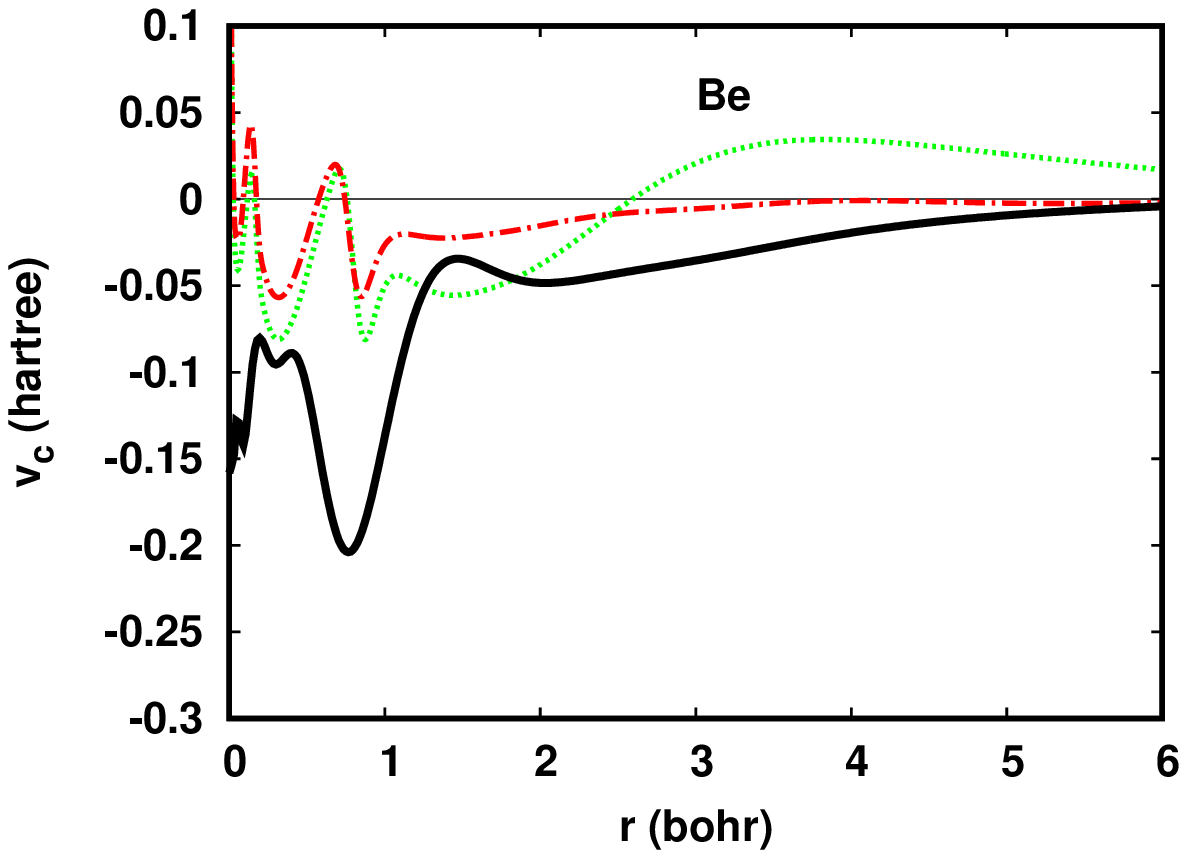}
\includegraphics[scale=0.45]{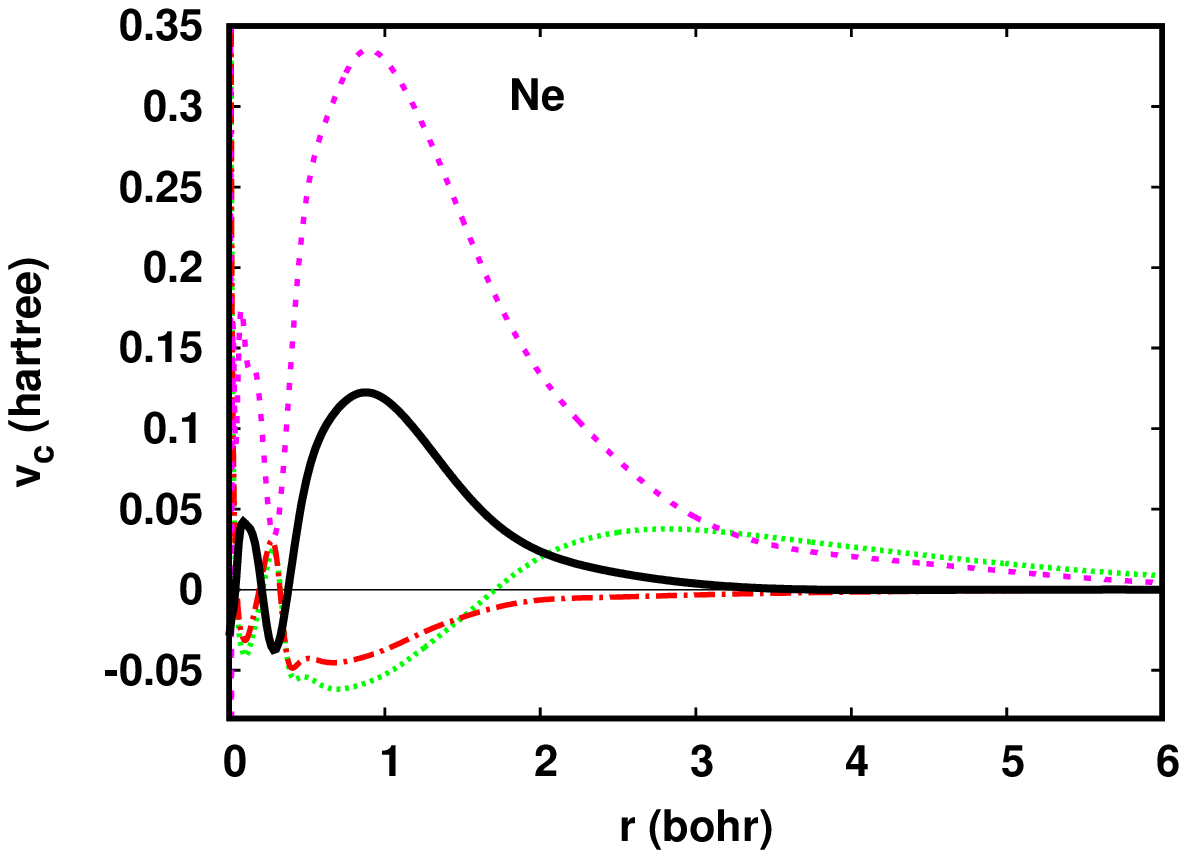}
\includegraphics[scale=0.45]{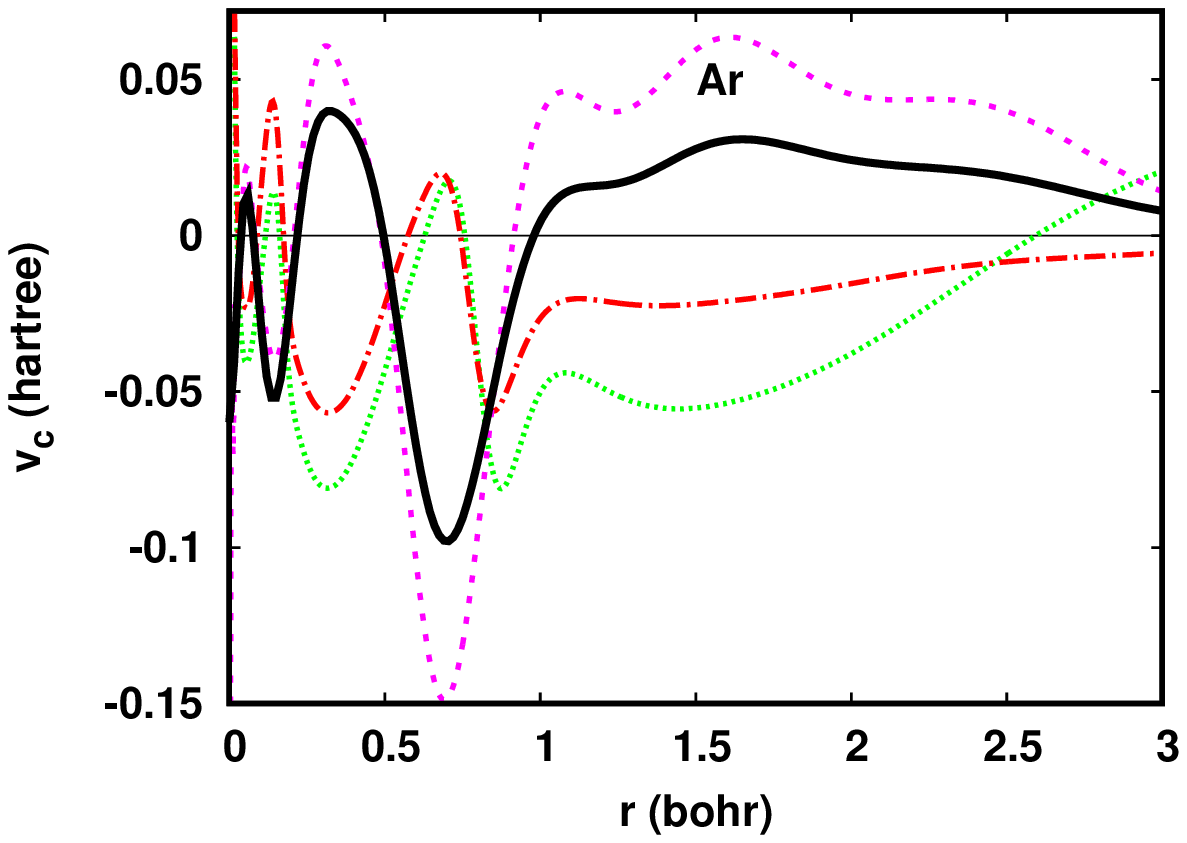}
\includegraphics[scale=0.45]{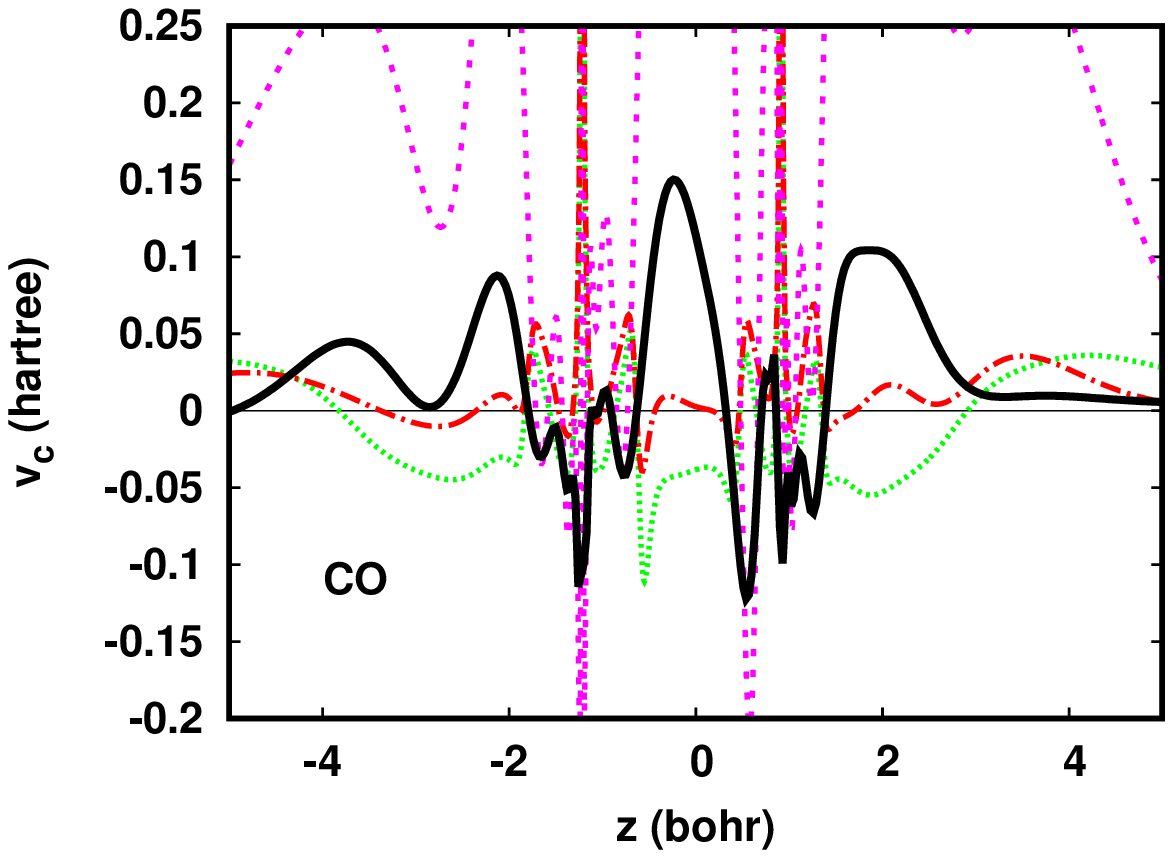}
\includegraphics[scale=0.45]{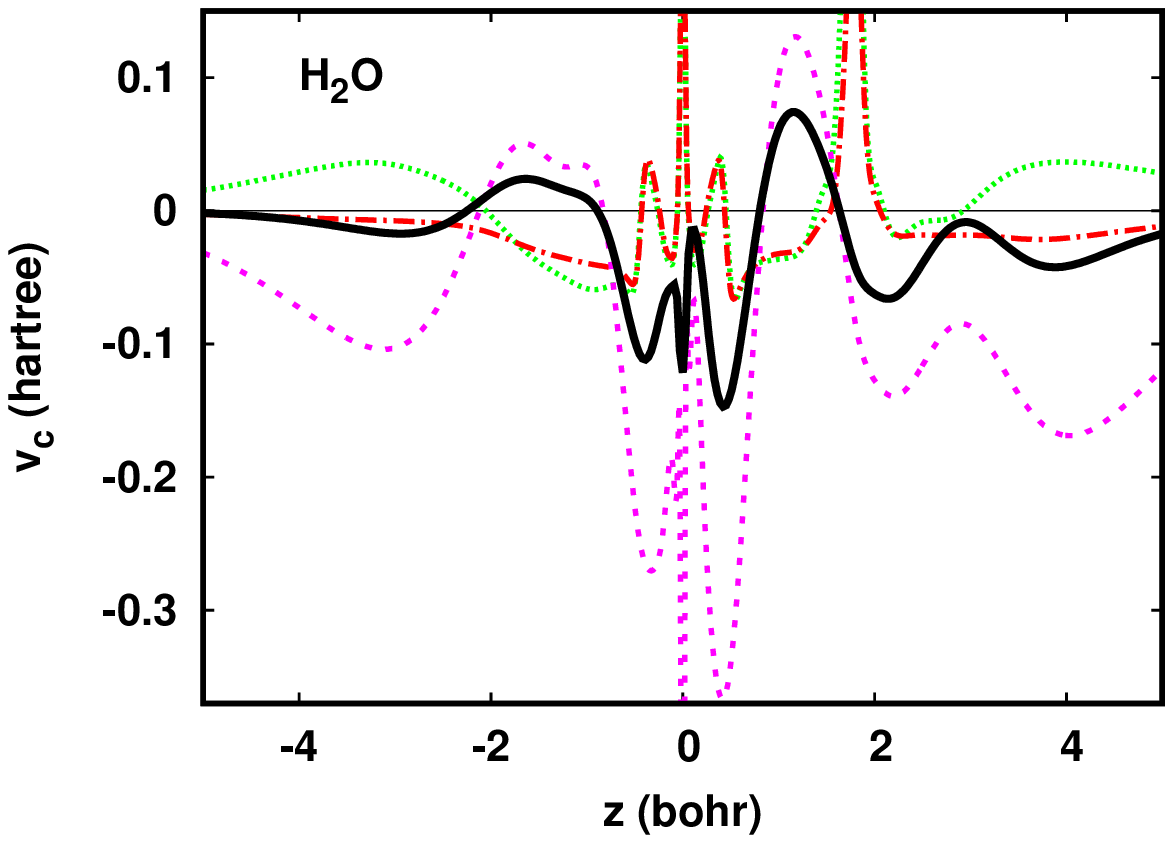}
\caption{Correlation potentials calculated with the RS-OEP2 method using the srPBE exchange-correlation density functional for the range-separation parameter $\mu=0.5$ bohr$^{-1}$, and for the limiting values $\mu = 0$ (standard KS PBE) and $\mu \to \infty$ (OEP-GL2). The reference potentials were calculated by KS inversion of CCSD(T) densities. For Be, the OEP-GL2 calculation is unstable.}
\label{fig:vc}
\end{figure*}
\begin{figure}
\includegraphics[scale=0.45]{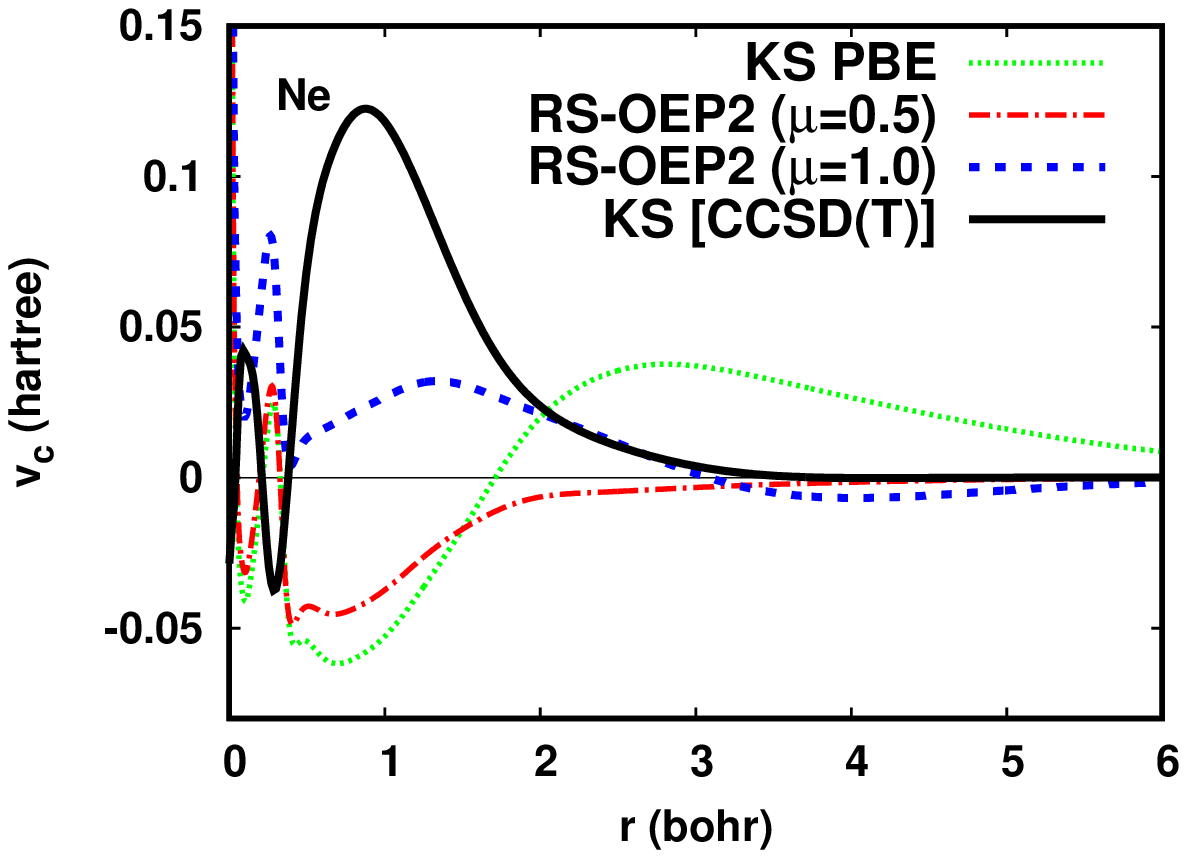}
\includegraphics[scale=0.45]{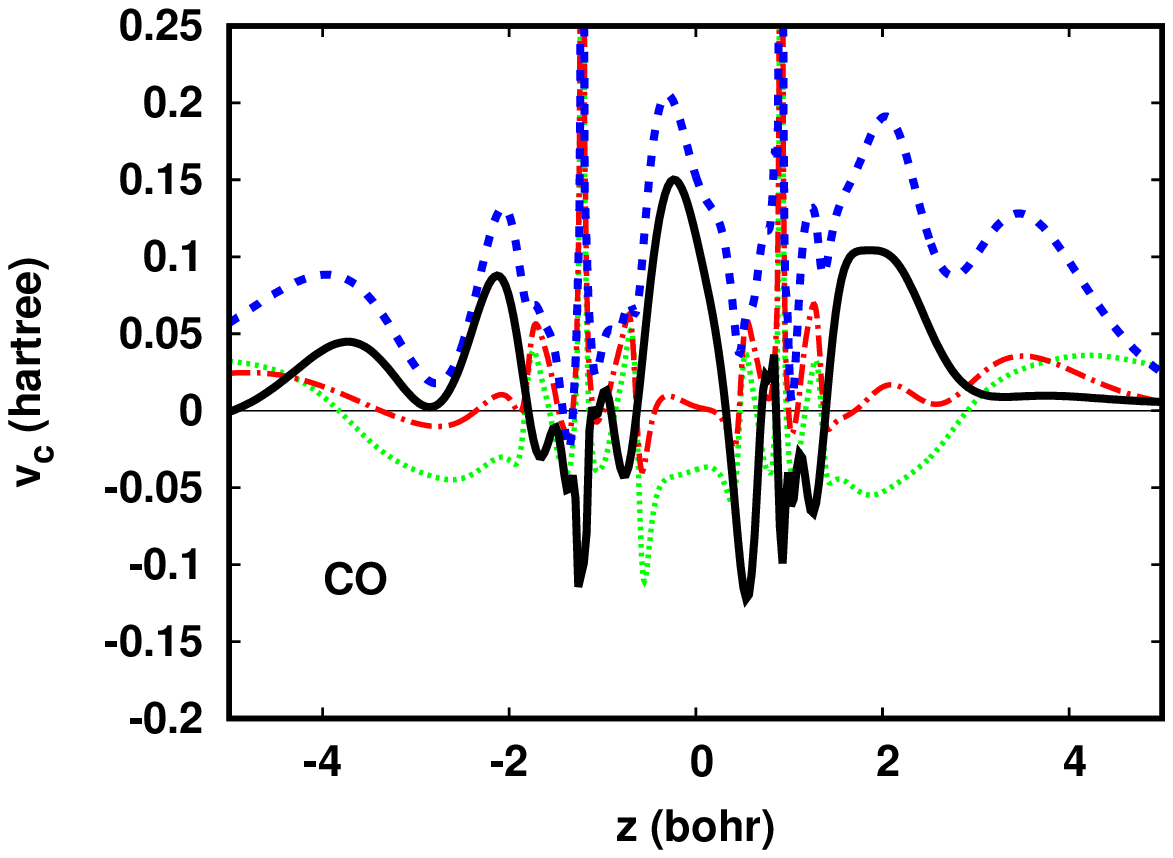}
\caption{Correlation potentials calculated with the RS-OEP2 method using the srPBE exchange-correlation density functional for the range-separation parameters $\mu=0.5$ bohr$^{-1}$ and $\mu=1$ bohr$^{-1}$, and for the limiting values $\mu = 0$ (standard KS PBE) and $\mu \to \infty$ (OEP-GL2). The reference potentials were calculated by KS inversion of CCSD(T) densities.}
\label{fig:vc2}
\end{figure}

We turn now our attention to the EAs. For the RSH+MP2 method, the EA is calculated similarly as the IP, i.e. as a finite-difference derivative of the total energy with respect to the electron number, as described in \Refs{CohMorYan-JCTC-09,SuYanMorXu-JPCA-14,SmiFraMusBukGraLupTou-JCP-16}. For the RS-OEP2 method, the EA is computed as a sum of LUMO energy and the derivative discontinuity coming from the long-range exchange and correlation OEP potentials, similarly as in \Ref{SmiFraMusBukGraLupTou-JCP-16}. These EAs are reported in \Fig{fig:ea} as a function of $\mu$. The reference CCSD(T) values (indicated by the top black horizontal lines) are calculated as the difference between the total energies of the $(N+1)$- and $N$-electron systems with the same basis sets. Note that the fact that we obtain positive $-$EA reflects the incompleteness of the basis set which artificially confines the additional electron. We also report the LUMO energies for the RSH and RS-OEP2 methods, as well as reference KS LUMO energies obtained by inversion of the KS equations using CCSD(T) densities as input (bottom black horizontal lines).

Again, at $\mu=0$, all methods reduces to KS PBE, which gives LUMO energies largely too low compared to the CCSD(T) reference. For $\mu \to \infty$, RSH reduces to standard HF, with a LUMO orbital energy which is a better estimate of $-$EA (the error ranges from about 0.01 to 1.2 eV). In the same limit, RSH+MP2 reduces to standard MP2, which gives EAs essentially identical to the CCSD(T) values, and RS-OEP2 reduces to OEP-GL2 which gives $-$EA values with an accuracy quite dependent on the system considered: i) in good agreement with CCSD(T) for Ne and Ar; ii) substantially too high for He and CO with errors of about 0.4 and 1.9 eV, respectively; iii) slightly to low for H$_2$O with an error of about 0.3 eV; and iv) incalculable for Be due to the instability. The LUMO energies given by OEP-GL2 are reasonable approximations to the accurate KS LUMO energies (except for the unstable case of Be). These LUMO energies are always negative and much lower than the opposite of EAs, and do not correspond to the addition of an electron but to a neutral excitation.

Around the common value of $\mu=0.5$ bohr$^{-1}$, RSH, RSH+MP2, and RS-OEP2 all give EAs reasonably close to the reference CCSD(T) values, and moreover the RS-OEP2 LUMO energies are also quite close to the accurate KS reference values. In order to investigate this more closely, we report the values of $-$EA for different values of $\mu$ in \Tab{tab:ea}. The largest errors are obtained with KS PBE, corresponding to the $\mu=0$ limit, with a MAE of 2.52 eV. Owing to the accuracy of standard MP2, RSH+MP2 gives decreasing errors with increasing $\mu$, with MAEs of 0.14, 0.09, and 0.05 eV for $\mu=0.5$, $\mu=1$ bohr$^{-1}$, and $\mu\to\infty$, respectively. By contrast, RS-OEP2 gives minimal MAEs of 0.35 and 0.38 eV for $\mu=0.5$ and $\mu=1$ bohr$^{-1}$, respectively, and a larger MAE of 0.55 eV in the limit $\mu\to\infty$.

As for the IPs, we thus conclude that self-consistency does not bring any improvement for the calculation of EAs in the RSH+MP2 approach. However, contrary to the RSH+MP2 method, the RS-OEP2 method gives a LUMO corresponding to a neutral excitation and its energy is a good approximation to the exact KS LUMO energy. This may be advantageous for calculating excitation energies.

\subsection{Exchange-correlation and correlation potentials}

\begin{figure*}
\includegraphics[scale=0.45]{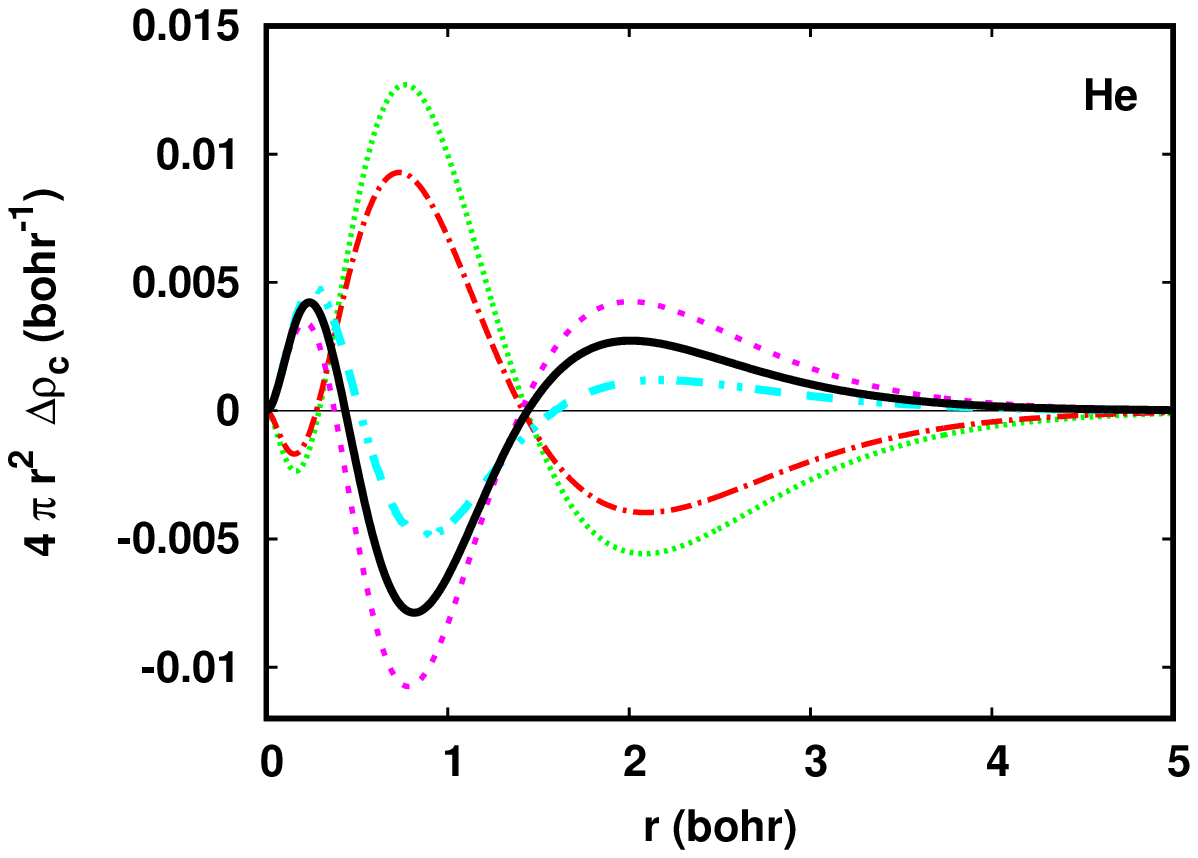}
\includegraphics[scale=0.45]{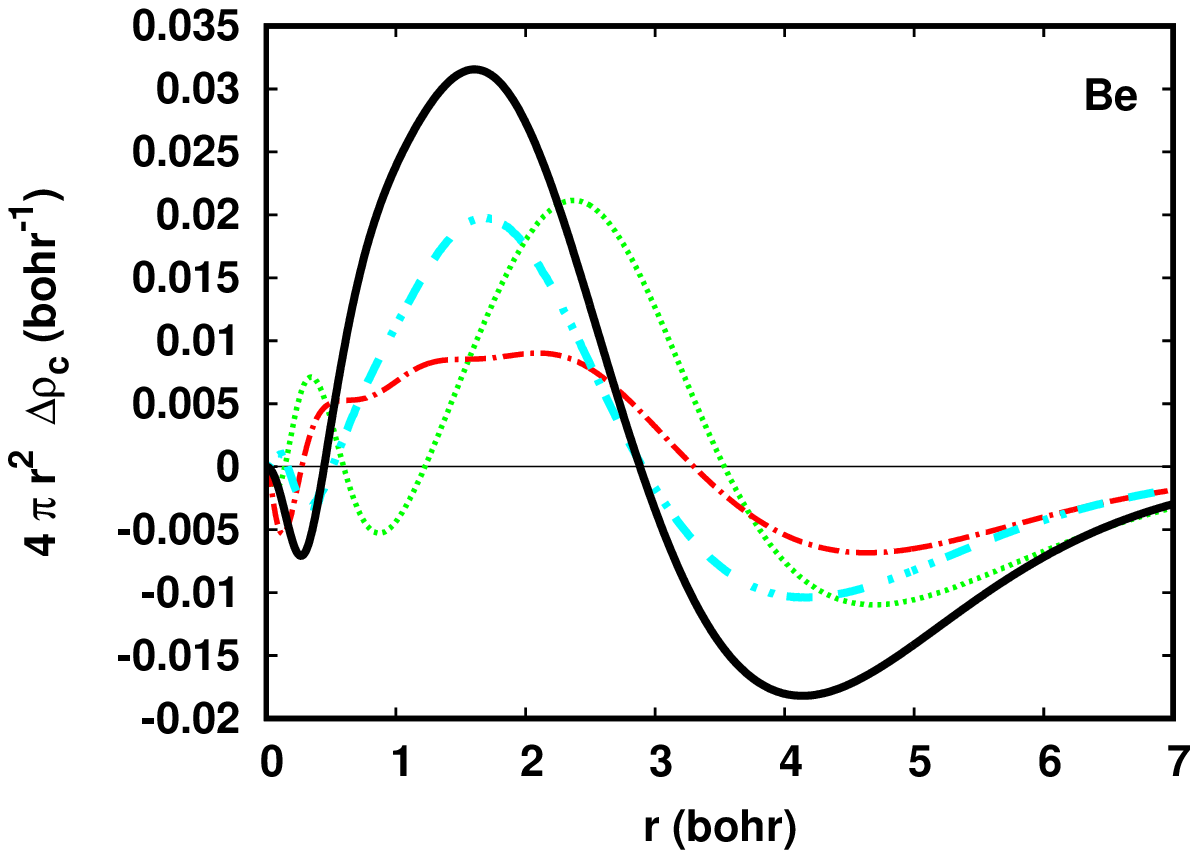}
\includegraphics[scale=0.45]{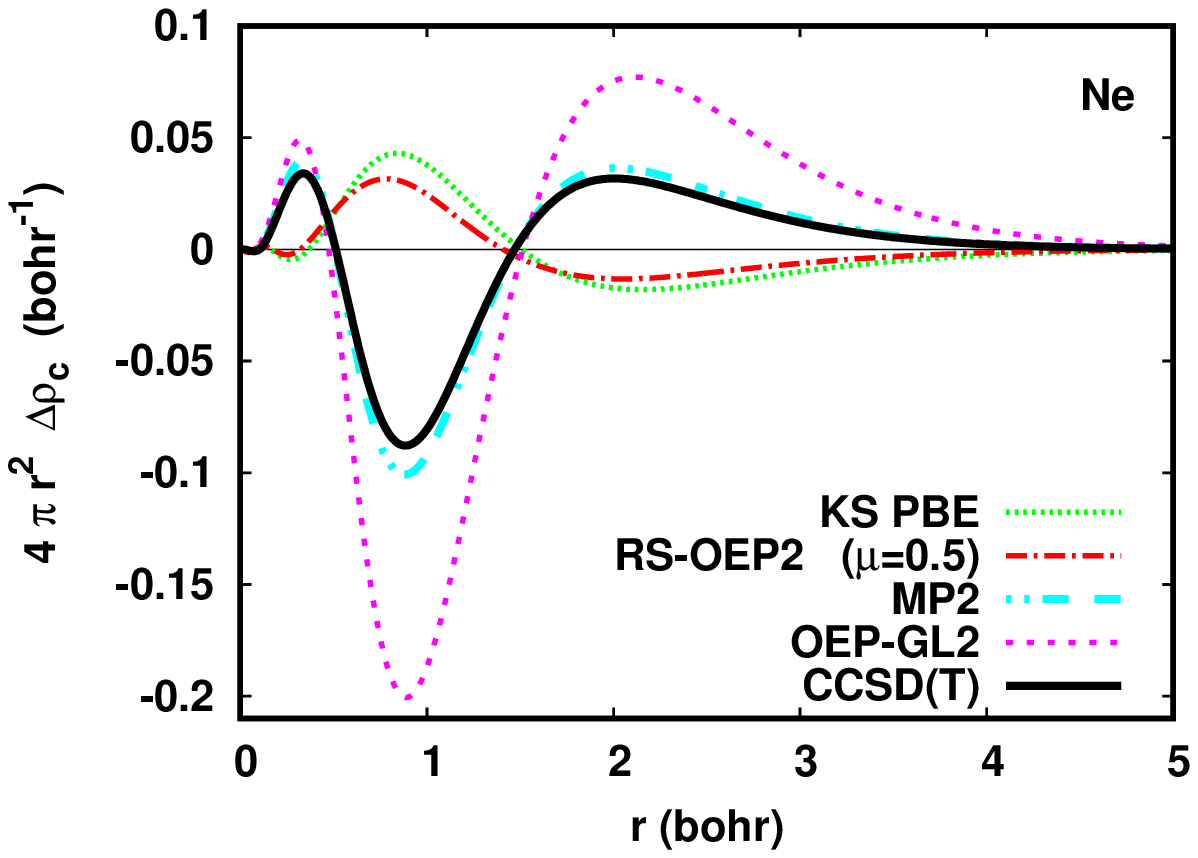}
\includegraphics[scale=0.45]{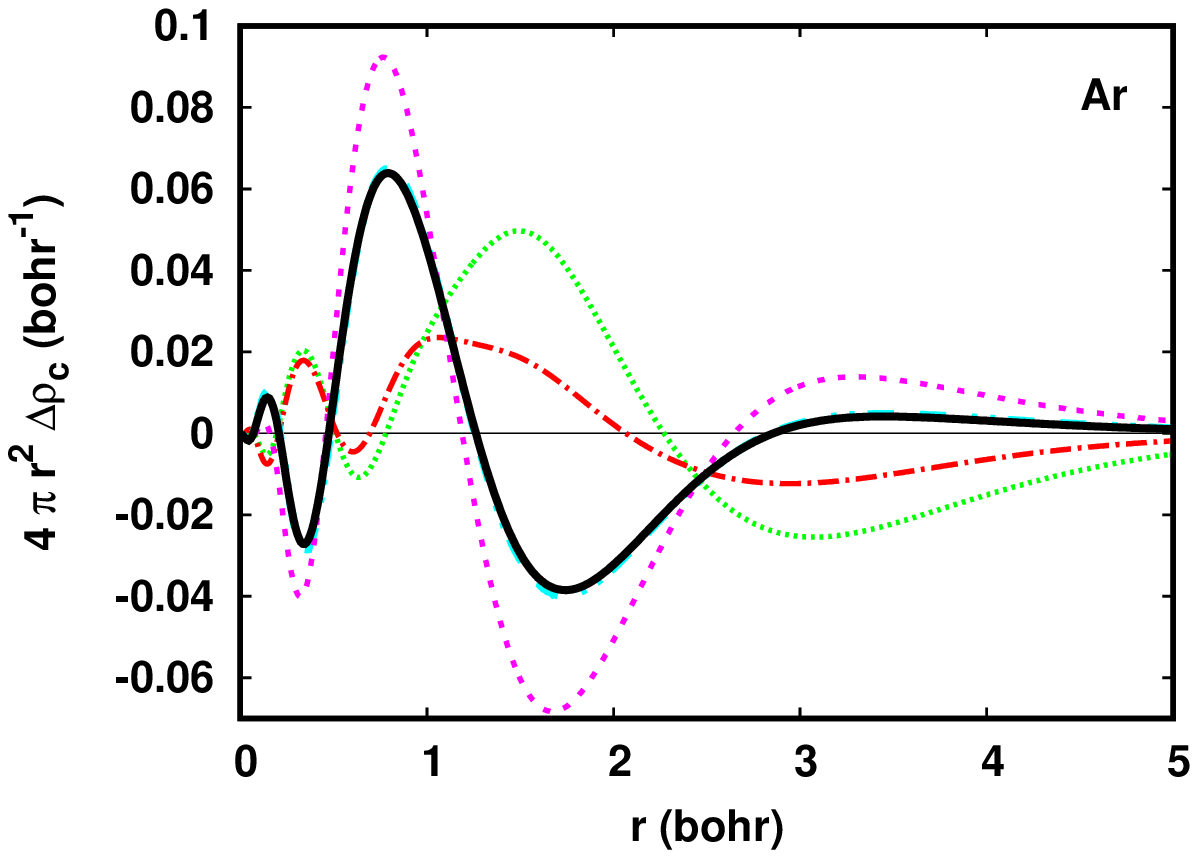}
\includegraphics[scale=0.45]{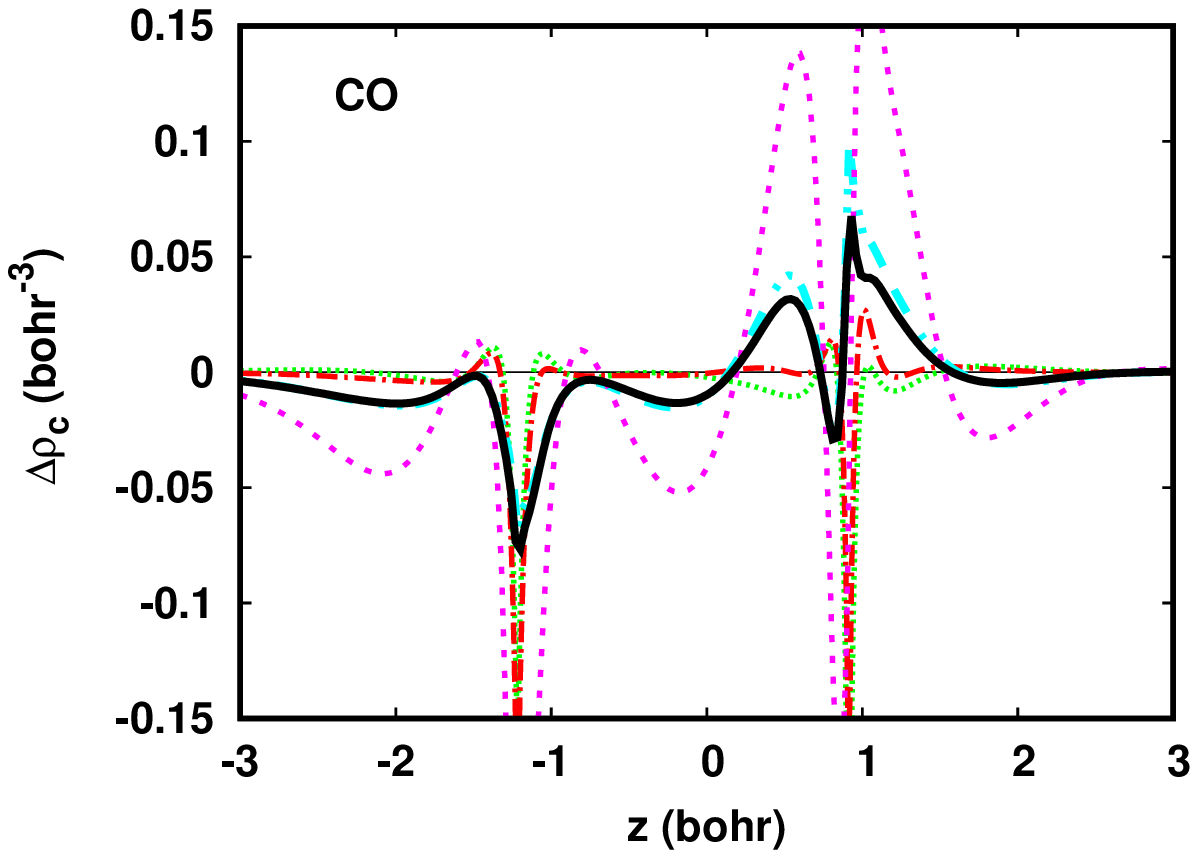}
\includegraphics[scale=0.45]{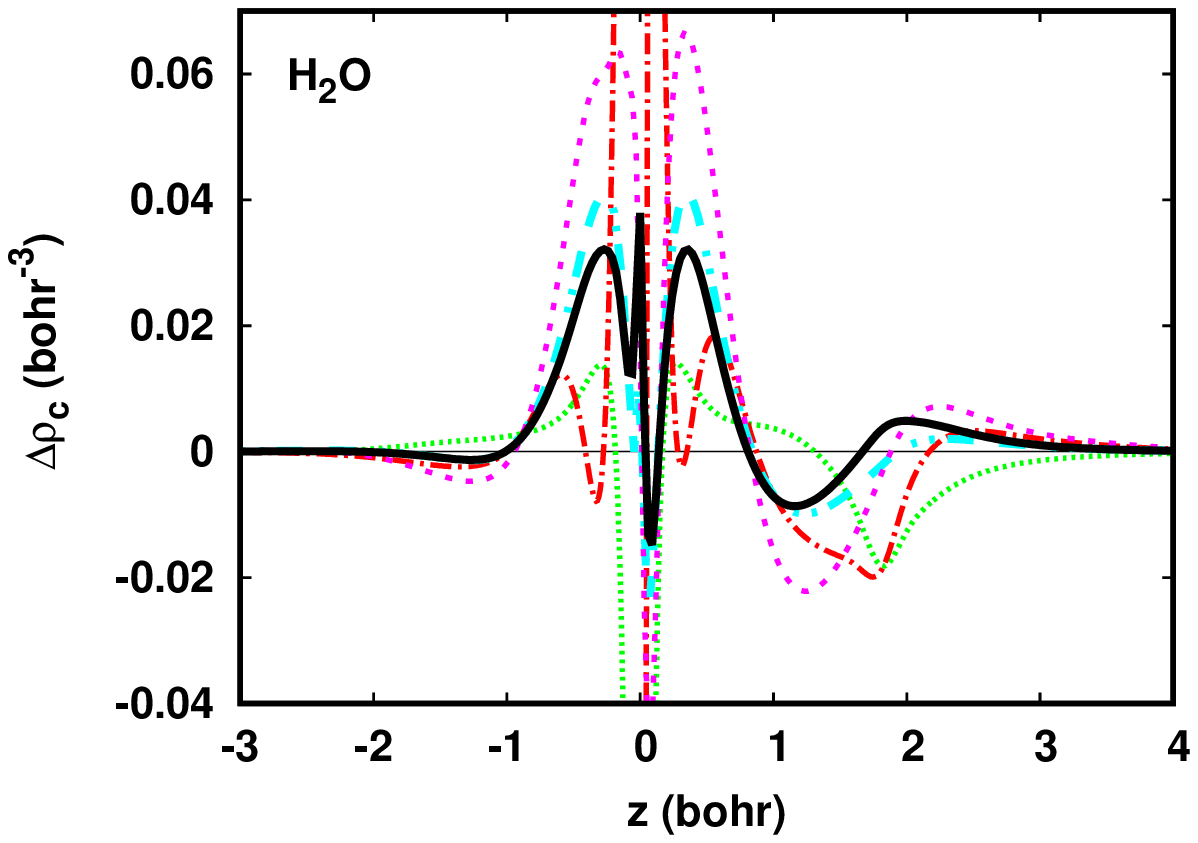}
\caption{Correlated densities calculated with the RS-OEP2 method using the srPBE exchange-correlation density functional for the range-separation parameter $\mu=0.5$ bohr$^{-1}$, and for the limiting values $\mu = 0$ (standard KS PBE) and $\mu \to \infty$ (OEP-GL2). For comparison, correlated densities calculated with standard MP2 are also shown. The reference correlated densities were calculated with CCSD(T). 
For Be, the OEP-GL2 calculation is unstable. For CO the correlated densities are calculated along the molecular axis,
and for H$_2$O along the HO bond.}
\label{fig:density}
\end{figure*}

One of the advantages of the RS-OEP2 method is that we have local exchange and correlation potentials, including both the long-range and short-range contributions, which is useful for analysis of the approximations and comparison to the exact KS scheme. We report in \Fig{fig:vxc} the RS-OEP2 exchange-correlation potentials for the commonly used value of the range-separation parameter $\mu=0.5$ bohr$^{-1}$, as well as for the limiting values $\mu = 0$ (standard KS PBE) and $\mu \to \infty$ (OEP-GL2). For comparison, we also report accurate exchange-correlation potentials obtained by KS inversion of CCSD(T) densities. 

First of all, we note that the PBE and RS-OEP2 exchange-correlation potentials diverge to minus infinity at the nuclei positions (see, e.g., the He atom). This is due to the inclusion of the density Laplacian term~\cite{BUK-2016} in the semilocal DFA part of the potential. The PBE exchange-correlation potentials only partially reproduce the shell structure of the accurate reference potentials. They are not negative enough in the valence region, and decay too fast at large distances which is the cause of the large underestimation of IPs by KS PBE. The OEP-GL2 method gives exchange-correlation potentials which are quite accurate for He and Ar, and to a lesser extent for Ne, but not negative enough for CO and too negative for H$_2$O. For Be, the OEP-GL2 calculation is unstable. The RS-OEP2 exchange-correlation potentials for $\mu=0.5$ bohr$^{-1}$ tend to be slightly not negative enough at small valence distances but have the correct $-1/r$ asymptotic behavior at large distances, as most clearly seen on atoms in \Fig{fig:vxc}. This feature is important for obtaining accurate IPs as well as accurate excitation energies within TDDFT~\cite{Hir-JCP-AC-2002}.  By contrast, the OEP-1DH method of \Ref{SmiFraMusBukGraLupTou-JCP-16} gives exchange-correlation potentials that tend to underestimate the long-range tail (see Fig. 5 of \Ref{SmiFraMusBukGraLupTou-JCP-16}).

The correlation potentials are shown in Fig.~\ref{fig:vc}. The PBE correlation potentials are out of phase with respect to the CCSD(T) reference potentials, i.e. minima of the PBE correlation potentials are often observed where the CCSD(T) correlation potentials have maxima. This qualitatively incorrect behavior of correlation potentials has been already observed for other GGAs~\cite{Grab-JCP-2011,BUK-2016} and meta-GGAs~\cite{Kanan-JCP-2013} functionals. The OEP-GL2 correlation potentials tend to be in phase with the CCSD(T) reference potentials, but they are largely overestimated, as previously observed in \Refs{Grab-JCP-2011,grabowski:2014:jcp}. In the core and in the valence regions, the RS-OEP2 correlation potentials are similar to the PBE ones. Farther from the nuclei, the RS-OEP2 correlation potentials tend to go more rapidly to zero than both the PBE and OEP-GL2 correlation potentials, and tend to correctly remove the unphysical long-range contributions of these potentials. We note in passing that, in the case of the Ar atom, the RS-OEP2 correlation potential is quite similar to the one obtained from recently developed adiabatic connection semi-local correlation (ACSC) functional (see Fig. 4 in \Ref{Const-ACSC-2019}).

In \Fig{fig:vc2} we show additionally, for Ne and CO, the RS-OEP2 correlation potentials for a larger value of the range-separation parameter, i.e. $\mu=1$ bohr$^{-1}$. In this case, the RS-OEP2 correlation potentials tend to be in better agreement with the reference potentials in the valence regions, which must come from the increased role of the long-range GL2 correlation term.

\subsection{Correlated densities}

In \Fig{fig:density} we report correlated densities calculated by the RS-OEP2 method for $\mu=0.5$ bohr$^{-1}$ and by the KS PBE, MP2, and OEP-GL2 methods. The reference correlated densities were calculated from the CCSD(T) relaxed density matrix. It have been shown that the analysis of the correlated densities can provide useful information in the development and testing of density functionals in KS DFT~\cite{Jankowski:2009:DRD,Jankowski:2010:DRD,Grab-JCP-2011,grabowski:2014:jcp,Smiga2014125,Gra-MP-2014,BUK-2016,SmiFraMusBukGraLupTou-JCP-16}. This quantity is defined as $\Delta \rho_\text{c} (\R) = \rho(\R) - \rho_\text{x}(\R)$ where $\rho(\R)$ is the total density calculated with the full exchange-correlation term and $\rho_\text{x}(\R)$ is the density calculated only at the exchange level. 

The KS PBE correlated densities are quite far the reference CCSD(T) correlated densities, being often out of phase with them, as was the case for the correlation potentials. This behavior was also reported for other density functionals in \Refs{Grab-JCP-2011,Gra-MP-2014}. The OEP-GL2 correlated densities have a qualitative behavior similar to the reference correlated densities but are largely overestimated. The RS-OEP2 correlated densities are overall quite similar to the KS PBE correlated densities.

\section{Conclusion}
\label{sec:conclusion}

We have extended the range-separated hybrid RSH+MP2 method to a fully self-consistent version using the OEP technique in which the orbitals are obtained from a local potential including the long-range HF exchange and MP2 correlation contributions. We have tested this approach, named RS-OEP2, using a short-range version of PBE exchange-correlation density functional, on a set of small closed-shell atoms and molecules. For the commonly used value of the range-separation parameter $\mu=0.5$ bohr$^{-1}$, while RS-OEP2 is a big improvement over KS PBE, it gives very similar total energies, IPs, and EAs than RSH+MP2. Thus, self-consistency itself does not seem to bring any improvement, at least for these systems and properties. One distinct feature of RS-OEP2 over RSH+MP2 is that it gives a LUMO energy which physically corresponds to a neutral excitation energy (and not to a EA) and which is a reasonably good approximation to the exact KS LUMO energy. Moreover, contrary to RSH+MP2, the RS-OEP2 method naturally gives local exchange-correlation potentials and densities. The RS-OEP2 exchange-correlation potentials are reasonable approximations to the exact KS exchange-correlation potentials and have the correct asymptotic behavior. However, a finer look at correlation potentials and correlated densities shows that RS-OEP2 barely improves over KS PBE for these quantities.

In future works, the RS-OEP2 method should be tested on more systems, including open-shell species where the benefices of self-consistency should be more important~\cite{PevHea-JCP-13}, and could be extended to the calculations of excitation energies in TDDFT. More importantly, the limited accuracy of the obtained correlation potentials and correlated densities may be overcome by introducing a fraction of HF exchange and MP2 correlation in the short-range part of electron-electron decomposition~\cite{KalTou-JCP-18,KalMusTou-JCP-19}. Another possible route of improvement is the development of a new type of short-range functional and potential which could be combined with state-of-art \textit{ab initio} DFT (OEP2) functionals\cite{grabowski:2007:ccpt2,Grab-JCP-2011,verma:044105,grabowski13,grabowski:2014:jcp,SCSIP}.

\section*{Acknowledgements}
S.S thanks the Polish National Science Center for the partial financial support under Grant No. 2016/21/D/ST4/00903.

\end{document}